\documentclass[epj,nopacs]{svjour}
\usepackage{amsmath,amsfonts,amssymb,bm}
\usepackage{graphicx}
\usepackage{hyperref}
\usepackage{color}
\usepackage{soul}
\usepackage{subfigure}
\usepackage{multirow}
\usepackage{textcomp}
\usepackage[USenglish]{babel}
\usepackage{slashed}
\usepackage[utf8]{inputenc}
\usepackage[bottom]{footmisc}
\usepackage{widetext}
\usepackage{float}
\begin{document}
\title{QCD phase diagram in a magnetized medium from the chiral symmetry perspective: The linear sigma model with quarks and the Nambu--Jona-Lasinio model effective descriptions}
\author{Alejandro Ayala\inst{1,2} \and Luis A. Hern\'andez\inst{3,2} \and Marcelo Loewe\inst{4,5,2} \and Cristian Villavicencio\inst{6}
}                     
%
%
\institute{Instituto de Ciencias Nucleares, Universidad Nacional Aut\'onoma de M\'exico, Apartado Postal 70-543, CdMx 04510, Mexico. \and Centre for Theoretical and Mathematical Physics, and Department of Physics, University of Cape Town, Rondebosch 7700, South Africa. \and Departamento de F\'isica, Universidad Aut\'onoma Metropolitana-Iztapalapa, Av. San Rafael Atlixco 186, CdMx 09340, Mexico. \and Instituto de F\'isica, Pontificia Universidad Cat\'olica de Chile, Casilla 306, Santiago, Chile. \and Centro Cient\'ifico Tecnol\'ogico de Valpara\'iso-CCTVAL,
Universidad T\'ecnica Federico Santa Mar\'ia, Casilla 110-V, Valpara\'iso, Chile. \and Centro de Ciencias Exactas and Departamento de Ciencias B\'asicas, Facultad de Ciencias, Universidad del B\'io-B\'io, Casilla 447, Chill\'an, Chile.}
\date{Received: date / Revised version: date}
%
\abstract{
We review the main features of the QCD phase diagram description, at finite temperature, baryon density and in the presence of a magnetic field, from the point of view of effective models, whose main ingredient is chiral symmetry. We concentrate our attention on two of these models: The linear sigma model with quarks and the Nambu--Jona-Lasinio model. We show that a main ingredient to understand the characteristics of the phase transitions is the inclusion of plasma screening effects that capture the physics of collective, long-wave modes, and thus describe a prime property of plasmas near transition lines, namely, long distance correlations. Inclusion of plasma screening makes possible to understand the inverse magnetic catalysis phenomenon even without the need to consider magnetic field-dependent coupling constants. Screening is also responsible for the emergence of a critical end point in the phase diagram even for small magnetic field strengths. Although versatile, the NJL model is also a more limited approach since, being a non-renormalizable model, a clear separation between pure vacuum and medium effects is not always possible. The model cannot describe inverse magnetic catalysis unless a magnetic field dependent coupling is included. The location of the critical end point strongly depends on the choice of the type of interaction and on the magnetic field dependence of the corresponding coupling. Overall, both models provide sensible tools to explore the properties of magnetized, strongly interacting matter. However, a cross talk among them as well as a consistent physical approach to determine the model parameters is much needed.
\PACS{
      {PACS-key}{discribing text of that key}   \and
      {PACS-key}{discribing text of that key}
     } 
} 
\maketitle
\section{Introduction}
\label{secI}

The study of the  different phases that strongly interacting matter can reach when varying control parameters, such as temperature, baryon and isospin density, etc., is nowadays one of the most active fields at the crossroads of nuclear and particle physics. The interest grew with the recent lattice QCD (LQCD) discovery of the inverse magnetic catalysis (IMC) phenomenon, whereby the critical temperature for the chiral symmetry restoration transition decreases as a function of the strength of another of the control parameters, an external magnetic field~\cite{Bali:2011qj,Bali:2012zg,Bali:2014kia}. This discovery has prompeted research on the properties of the so called {\it magnetized QCD phase diagram}. In addition, the recent successful detection of gravitational waves~\cite{TheLIGOScientific:2017qsa} from a binary neutron star merger, has kicked off the era of multi-messenger physics to study strongly interacting systems. The information obtained by studying these signals can be combined with information from heavy-ion experiments to get a clearer picture of the properties of hadronic matter at high densities, temperatures and in the presence of magnetic fields.

Magnetic fields contribute to the properties of a large variety of physical systems including heavy-ion collisions \cite{2008NuPhA.803..227K,MCLERRAN2014184}, the interior of compact astrophysical objects~\cite{Duncan:1992hi,2018arXiv180305716E,Ayala:2018kie} and even the early universe~\cite{VACHASPATI1991258,Navarro:2010eu,Sanchez:2006tt}. It has been estimated that the magnetic field strength $|eB|$ in peripheral heavy-ion collisions reaches values equivalent to a few times the pion mass squared, both at RHIC and at the LHC~\cite{2009IJMPA..24.5925S}. The effects of such magnetic fields cannot be overlooked in a complete description of these systems and its understanding contributes, at a fundamental level, to a better characterization of the properties of QCD matter~\cite{PhysRevC.91.064902,PhysRevD.90.085011,PhysRevD.92.096011,PhysRevD.92.016006,PhysRevD.91.016007,PhysRevD.91.016002,PhysRevD.89.116017,PhysRevD.89.016004,Ayala:2018wux}.

The QCD phase diagram consists of an idealized picture, where the transition lines correspond to the boundaries between different phases of strongly interacting matter. Close to the phase boundaries, the relevant quark species are the light quarks $u$, $d$ and $s$. A complete description, accounting for the abundance of these species, should in principle be given in terms of the chemical potentials associated to each of these quarks. Nevertheless, under the requirements of beta equilibrium and charge neutrality, these chemical potentials are not independent from each other. Therefore, out of the three chemical potentials only one is independent. Any one of them can be chosen and the usual choice is the baryon chemical potential $\mu_B$, related to the quark chemical potential $\mu$ by $\mu=\mu_B/3$.

The shape of the possible transition lines has been conjectured since long ago using several general arguments. One of them goes as follows: when  nuclear matter is heated up, resonance hadronic states become excited. The density of states $\rho$ increases exponentially as a function of the resonance mass $m$, namely $\rho(m)\propto\exp\left\{ m/T^H  \right\}$, where $T^H\simeq 0.19$ GeV. This density of states competes with the Boltzmann phase space occupation factor $\rho_B\propto\exp\left\{ -m/T \right\}$, namely
\begin{eqnarray}
   \rho(m)\rho_B(m)=\exp\left\{\frac{m}{T^H}-\frac{m}{T}\right\},
\label{competes}
\end{eqnarray}
such that when $T>T^H$, the integration over $m$ becomes singular.
$T^H$ plays the role of a limiting temperature known as the Hagedorn temperature above which the hadronic description breaks down~\cite{Hagedorn:1976ef}. Applying a similar argument, we can also estimate the critical line at finite $\mu_B$. The density of baryon states $\rho(m_B)\propto\exp\left\{m_B/T^H\right\}$, where $m_B$ is the typical baryon mass (of order 1 GeV) should be balanced by the Boltzmann factor
\begin{eqnarray}
\exp\left\{-(m_B-\mu_B)/T\right\}.
\end{eqnarray} Therefore, the line describing the relation between the limiting temperature and baryon chemical potential becomes
\begin{eqnarray}
   T=\left(1-\frac{\mu_B}{m_B}\right)T^H.
\label{limitingbaryon}
\end{eqnarray}

Another qualitative argument to draw the transition lines can be provided from QCD in the large number of colors ($N_c$) limit, while keeping the number of flavors ($N_f$) fixed~\cite{McLerran:2007qj}. When the quark
chemical potential $\mu\gtrsim\Lambda_{QCD}$, baryons form
a dense phase where the pressure is ${\mathcal{O}}(N_c)$. This dense phase is still confined but chirally symmetric, and is called the {\it quarkyonic} phase. The chiral phase transition happens in this phase and if a critical end point (CEP) exists then the deconfining and chiral transitions split from one another at that point.

LQCD has been successfully applied to find the chiral/deconfining transition temperature for $\mu_B=0$. The result is that a crossover occurs at a pseudocritical temperature $T_c(\mu_B=0)\simeq 155$ MeV~\cite{Bazavov:2018mes}. Unfortunately, LQCD calculations cannot be used to determine the position of a possible CEP, due to the severe sign problem. Nevertheless, recent results employing the Taylor series expansion around $\mu_B=0$ or the extrapolation from imaginary to real $\mu_B$  values, show that the CEP is not to be found for $\mu_B/T\leq 2$ and $145\leq T\leq 155$ MeV~\cite{Sharma:2017jwb}. A more recent bound disfavors the existence of a CEP for $\mu_B/T \leq 2$ and $T/T_c(\mu_B = 0) > 0.9$~\cite{Bazavov:2017dus}.

The statistical model~\cite{Andronic:2005yp} can be used to map out the chemical freeze-out curve in relativistic heavy-ion collisions, as a function of the collision energy, from fits to particle ratios. These fits provide the temperature and baryon chemical potential at chemical freeze-out, namely, when particle scattering does not change the abundance of the different hadron species. Remarkably, this curve coincides, within uncertainties, with the LQCD results for the transition curve between the confined and the deconfined phases. It is difficult to believe that this coincidence happens just by chance. It has been argued that when the phase transition line is crossed, multiparticle scattering of Goldstone bosons drives baryons rapidly into equilibrium~\cite{Braun_Munzinger_2004}. This effect may provide an explanation for the observation that the chemical freeze-out line reaches the phase
boundary. An outstanding question is then: how, if at all, the presence of an external magnetic field modifies the transition lines and in particular the location of a possible CEP? In spite of its limitations to explore the phase diagram by and large, LQCD calculations show that for very strong magnetic
fields, IMC prevails and the
phase transition becomes first order at asymptotically
large values of the magnetic field for vanishing quark
chemical potential~\cite{Endrodi:2015oba}. A similar behavior is obtained in
the Nambu--Jona-Lasinio (NJL) model if one includes a magnetic
field dependence of the critical temperature~\cite{Costa:2015bza,Ferreira:2015jra}, an idea first put forward in  Ref.~\cite{Farias:2014eca}. In order to answer the above question, it is then  necessary to resort to the use of theoretical tools that account for the two main features of QCD relevant for the description of the phase structure of QCD namely, chiral symmetry and confinement. 

In this work we review the state of the art of research that makes use of effective models than aim to address the above question. We pay particular attention to the Linear Sigma Model with quarks (LSMq), also known in the literature as the quark meson model and to the NJL model. For the former, we show that when the meson sector is treated as dynamical, {\it i.e.}, meson fluctuations contribute to the thermo-magnetic properties of the system, it is possible to understand IMC and thus to extend these treatment to the description of the phase structure of QCD in the presence of magnetic fields from the chiral symmetry perspective. The work is organized as follows: In Sec~\ref{secII} we describe in great detail the elements that make up the LSMq in the presence of a magnetic field. These include the computation of the effective potential at one-loop level both for bosons and quarks, treating all particles as full quantum fields and thus allowing for their fluctuations. As we show, the boson contribution needs to be supplemented by the inclusion of the plasma screening effects encoded in the calculation of the ring diagrams contribution to the effective potential. In order to avoid the shift of the tree-level vacuum position and curvature introduced by the vacuum one-loop corrections, we also discuss the way these corrections can be absorbed into the vacuum stabilization conditions. We also discuss how the model parameters can be fixed invoking physical conditions near the transition line at finite temperature as well as from vacuum. The inclusion of magnetic field effects is made using Schwinger's proper-time formalism in the weak field limit, as appropriate for instance to describe the conditions during a peripheral heavy-ion collision. All along this section we work in the strict chiral limit where pions are massless in vacuum. Since in this limit, the thermo-magnetic corrections to the couplings turn out to be inversely proportional to the particle masses, their inclusion requires a special treatment that we omit in this work but plan to discuss in a future one. We provide a thorough analysis of the LSMq description of IMC and of the properties of the magnetized QCD phase diagram from the point of view of chiral symmetry emphasizing the crucial ingredient introduced by the plasma screening effects. In Sec.~\ref{secIII} we discuss how the NJL model can be used in the presence of a magnetic field to study the magnetized phase diagram. We emphasize that from the perspective of this model, a   better  understanding  of the evolution  of the CEP requires inclusion of crucial ingredients such as confinement , IMC and the vector-current interaction. Finally in Sec.~\ref{concl} we make concluding remarks and provide a prospective of the kind of studies that can be carried out in the near future to achieve a clearer picture on the subject. We point out that a recent review on some of these aspects can be found in Ref.~\cite{Andersen:2021lnk} and a thorough review of the known aspects of IMC is provided in Ref.~\cite{Bandyopadhyay:2020zte}.

\section{The Linear Sigma Model with quarks}\label{secII}

Effective models are useful proxies to help identify the
main characteristics of the QCD phase diagram. While no single model can be used to describe the whole extent of the phase diagram, they can be employed to explore different regions
with varying degrees of sophistication and inclusion
of effective degrees of freedom. For instance, Ref.~\cite{Krein:2021sco} works with the LSMq at finite temperature in the presence of strong magnetic fields to study the quark condensate time evolution using Langevin dynamics. Reference~\cite{Kawaguchi:2021nsa} explores the possible existence of a new CEP using a generic chiral model away from the chiral limit and in the presence of a weak magnetic field. Also, in Refs.~\cite{Herbst:2010rf,Carlomagno:2016bpu,Carlomagno:2019wlh}
a Polyakov-quark-meson model,
is employed to map the deconfinement and chiral symmetry
restoration transitions, finding that the crossover region for one and the other coincides within a band representing the width of the susceptibility
peak and that the width of such band shrinks as $\mu_B$ increases up to the region where a CEP at low temperature
values is found. Reference~\cite{Tawfik:2017cdx} studies magnetic properties of QCD matter in a thermal and dense medium, from both the chiral and deconfinement aspects, in an $SU(3)$ Polyakov linear-sigma model. The authors find that when the mean field approximation in the LSM is supplemented with the Polyakov loop, the model describes several properties of magnetizes QCD as found by LQCD, in particular IMC.

Given that LQCD calculations, extended to small but finite values of $\mu_B$, find coincident transition lines for the deconfinement and chiral symmetry restoration transitions, it should be possible to explore
the phase diagram emphasizing independently either
the deconfinement or the chiral aspects of the transition. 

An effective model that accounts for the latter is provided by the LSMq. The Lagrangian, including the coupling of charged particles to an external magnetic field, is given by 
\begin{eqnarray}
   \!\!\!\!\!\!\mathcal{L}&=&\frac{1}{2}(\partial_\mu \sigma)^2  + \frac{1}{2}(D_\mu \vec{\pi})^2 + \frac{a^2}{2} (\sigma^2 + \vec{\pi}^2)\nonumber\\
   &-& \frac{\lambda}{4} (\sigma^2 + \vec{\pi}^2)^2 + i \bar{\psi} \gamma^\mu D_\mu\psi -g\bar{\psi} (\sigma + i \gamma_5 \vec{\tau} \cdot \vec{\pi} )\psi ,
\label{lagrangian}
\end{eqnarray}
where $ q$ is an $SU(2)$ isospin doublet of quarks,
\begin{eqnarray}
 \vec{\pi}=(\pi_1, \pi_2, \pi_3 ),
\end{eqnarray} is an isospin triplet and $\sigma$ is an isospin singlet, with
\begin{eqnarray}
   D_{\mu}=\partial_{\mu}+iq_{f,b}A_{\mu},
\label{dcovariant}
\end{eqnarray}
being the covariant derivative with $q_{b,f}$ being the boson or fermion electric charge. $A^\mu$ is the vector potential corresponding to an external magnetic field directed along the $\hat{z}$ axis. In the symmetric gauge it is given by
\begin{eqnarray}
   A^\mu=\frac{B}{2}(0,-y,x,0).
\label{vecpot}
\end{eqnarray}
$A^\mu$ satisfies the gauge condition $\partial_\mu A^\mu=0$. The gauge field couples only to quarks and to the charged pion combinations, namely
\begin{eqnarray}
   \pi_\pm=\frac{1}{\sqrt{2}}\left(\pi_1\pm i\pi_2\right).
\end{eqnarray}
The neutral pion is taken as the third component of the pion isovector, $\pi^0=\pi_3$. The gauge field is considered as classical and thus there are no loops involving the propagator of the gauge field in internal lines. The squared mass parameter $a^2$ and the self-coupling $\lambda$ and $g$ are taken to be positive and, for the purpose of describing the chiral phase transition at finite $T$ and $\mu_B$, they need to be determined from conditions close to the phase boundary, and not only from vacuum conditions.
\begin{figure*}[t]
    \begin{center}
    \includegraphics[scale=0.8]{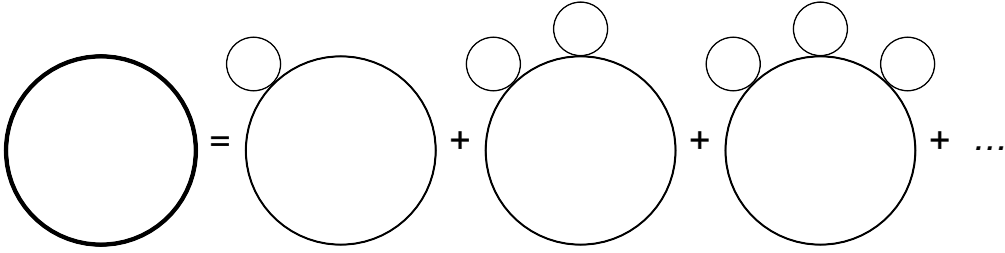}
    \end{center}
    \caption{Illustration of the contribution of the resummation of ring diagrams to the effective potential.}
    \label{rings}
\end{figure*}

To allow for spontaneous symmetry breaking, we let the $\sigma$ field to develop a vacuum expectation value $v$
\begin{eqnarray} \sigma \rightarrow \sigma + v.
\label{shift}
\end{eqnarray}
This vacuum expectation value can be identified with the order parameter of the theory. After this shift, the Lagrangian can be rewritten as
\begin{eqnarray}
{\mathcal{L}}&=& -\frac{1}{2}[\sigma(\partial_{\mu}+iqA_{\mu})^{2}\sigma]-\frac{1}
   {2}\left(3\lambda v^{2}-a^{2} \right)\sigma^{2}\nonumber\\
   &-&\frac{1}{2}[\vec{\pi}(\partial_{\mu}+iq_bA_{\mu})^{2}\vec{\pi}]-\frac{1}{2}\left(\lambda v^{2}- a^2 \right)\vec{\pi}^{2}\nonumber\\
   &+&\frac{a^{2}}{2}v^{2} -\frac{\lambda}{4}v^{4} + i \bar{\psi} \gamma^\mu D_\mu\psi
  -gv \bar{\psi}\psi + {\mathcal{L}}_{I}^b + {\mathcal{L}}_{I}^f,\nonumber\\
  \label{lagranreal}
\end{eqnarray}
where ${\mathcal{L}}_{I}^b$ and  ${\mathcal{L}}_{I}^f$ are given by
\begin{eqnarray}
  {\mathcal{L}}_{I}^b&=&-\frac{\lambda}{4}\Big[(\sigma^2 + (\pi^0)^2)^2+ 4\pi^+\pi^-(\sigma^2 + (\pi^0)^2 + \pi^+\pi^-)\Big],\nonumber\\
  {\mathcal{L}}_{I}^f&=&-g\bar{\psi} (\sigma + i \gamma_5 \vec{\tau} \cdot \vec{\pi} )\psi.
  \label{lagranint}
\end{eqnarray}
The terms given in Eq.~(\ref{lagranint}) describe the interactions among the fields $\sigma$, $\vec{\pi}$ and $ q$, after symmetry breaking. From Eq.~(\ref{lagranreal}) one can see that the $\sigma$, the three pions and the quarks have masses given, respectively, by
\begin{eqnarray}
  m^{2}_{\sigma}&=&3  \lambda v^{2}-a^{2},\nonumber\\
  m^{2}_{\pi}&=&\lambda v^{2}-a^{2}, \nonumber\\
  m_{f}&=& gv.
\label{masses}
\end{eqnarray}

A common, albeit limited, approximation is to work in the large number of colors case, where mesons are only included at tree-level, while fermions are considered as the true quantum particles in the system~\cite{Andersen:2021lnk}. In this approximation, fermion fluctuations experience the effects of a mean field provided by mesons. Part of the reason to consider this approximation is that, according to Eq.~(\ref{masses}), if mesons become true quantum fields, their masses are subject to change as the order parameter $v$ changes from its value $v_0=\sqrt{a^2/\lambda}$ given by the minimum of the tree-level potential, to its value $v_0=0$ at the restored phase when including thermal effects. During this transit, the meson square masses can become zero or even negative. 

This apparent drawback is in fact the key to properly account for meson fluctuations in a consistent manner. It is well known that when in a medium boson masses are small and their thermal corrections of the same order as the original masses, the latter need to be resummed. The naive perturbative expansion breaks down and the next to the one-loop contribution to thermodynamical quantities, in the limit when the number of bosons is large~\cite{Dolan:1973qd}, is the correction introduced by the {\it ring diagrams}. The name stems from the kind of Feynman diagrams that are resummed. These are illustrated in Fig.\ref{rings}.
Implementing the resummation program is equivalent to account for the plasma screening effects, whereby the coherent effect of long wave-length fluctuations prevent the appearance of infrared divergences. To illustrate the effect, consider the {\it effective potential}, which represents the quantity whose negative, upon integration over the system's volume and exponentiation, gives rise to the partition function. For a single boson species, the expressions at finite temperature $T$ for the one-loop effective potential and for the ring diagram contribution are given by~\cite{Bellac:2011kqa,Kapusta:1989tk}
\begin{eqnarray}
V_{b}^{(1)}=\frac{T}{2}\sum_n\int \frac{d^3k}{(2\pi)^{3}}\ln\left[\Delta_{b}(i\omega_n,{\mathbf {k}})\right]^{-1},
\label{eff1loop}
\end{eqnarray}
\begin{eqnarray}
V_{b}^{(ring)}=\frac{T}{2}\sum_n\int \frac{d^3k}{(2\pi)^{3}}\ln\left[1+\Pi_b(i\omega_n,{\mathbf {k}})\Delta_{b}(i\omega_n,{\mathbf {k}})\right],\nonumber \\
\label{effringprin}
\end{eqnarray}
respectively, where
\begin{eqnarray}
\Delta_b(i\omega_n,{\mathbf {k}})=\frac{1}{\omega_n^2 + {\mathbf {k}}^2 + m_b^2},
\label{bosprop}
\end{eqnarray}
is the Matsubara propagator and $\Pi_b$ the boson self-energy, for the time being not including magnetic field effects. Also, $\omega_n=2n\pi T$ are boson Matsubara frequencies. Notice that Eq.~(\ref{effringprin}) can also be written as
\begin{eqnarray}
V_{b}^{(ring)}&=&\frac{T}{2}\sum_n\int \frac{d^3k}{(2\pi)^{3}}\ln\left[\left(\Delta_{b}^{-1}(i\omega_n,{\mathbf {k}})+\Pi_b(i\omega_n,{\mathbf {k}})\right)\right.\nonumber\\
&\times&\left.\left(\Delta_{b}(i\omega_n,{\mathbf {k}})\right)\right]\nonumber\\
&=&\frac{T}{2}\sum_n\int \frac{d^3k}{(2\pi)^{3}}\ln\left[\Delta_{b}(i\omega_n,{\mathbf {k}})
\right]\nonumber\\
&+&
\frac{T}{2}\sum_n\int \frac{d^3k}{(2\pi)^{3}}\ln\left[\Delta_{b}^{-1}(i\omega_n,{\mathbf {k}})+\Pi_b(i\omega_n,{\mathbf {k}})\right].\nonumber\\
\label{effring-2}
\end{eqnarray}
Thus, by adding Eqs.~(\ref{eff1loop}) and~(\ref{effring-2}), we obtain
\begin{eqnarray}
V_{b}^{(1)}+V_{b}^{(ring)}&=&\frac{T}{2}\sum_n\int \frac{d^3k}{(2\pi)^{3}}\ln\left[\Delta_{b}^{-1}(i\omega_n,{\mathbf {k}})\right. \nonumber\\
&+&\left.\Pi_b(i\omega_n,{\mathbf {k}})\right],
\label{Veffsum}
\end{eqnarray}
which reveals that the boson mass squared $m_b^2$ is effectively replaced by the combination $m_b^2+\Pi_b$. Therefore, although the tree-level boson squared mass can become zero or even negative, the contribution to the mass coming from the boson self-energy makes the thermal square mass to be positive definite. In practice, in order to obtain analytical results, the correction introduced by the ring diagrams is computed in the high-temperature limit. The effect is to replace odd powers of the boson mass $m_b$, that appear in the large temperature expansion of the effective potential at one-loop, by $\sqrt{m_b^2+\Pi_b}$~\cite{Ayala:2014mla}.

\subsection{One-loop boson contribution}\label{secII-1}

We now turn to finding the boson contribution to the effective potential for the LSMq in the presence of a magnetic field~\cite{Ayala:2014gwa}. The Matsubara propagator for a boson with electric charge $q_b$ can now be written in terms of Schwinger's proper time representation as
\begin{eqnarray}
&\Delta_b&\!\!(i\omega_n,{\mathbf {k}};|q_bB|)=\int_0^\infty \frac{ds}{\cosh |q_bB|s}\nonumber \\
&\times& \exp\left\{
-s\left(\omega_n^2+k_3^2+k_\perp^2\frac{\tanh|q_bB|s}{|q_bB|s}+m_b^2\right)
\right\}\!.\nonumber\\
\label{bospropSch}
\end{eqnarray}
Therefore, the one-loop contribution to the effective potential Eq.~(\ref{eff1loop}) becomes in the presence of the magnetic field
\begin{eqnarray}
V_{b}^{(1;B)}
=\frac{T}{2}\sum_n\int \frac{d^3k}{(2\pi)^{3}}\ln\left[\Delta_{b}(i\omega_n,{\mathbf {k}};|q_bB|)\right]^{-1}.
\label{eff1loopB}
\end{eqnarray}
Using that
\begin{eqnarray}
&&\ln\left[\Delta_{b}(i\omega_n,{\mathbf {k}};|q_bB|)\right]^{-1}\nonumber\\
&=&\int dm_b^2\left(
\frac{d}{dm_b^2}\ln\left[\Delta_{b}(i\omega_n,{\mathbf {k}};|q_bB|)\right]^{-1}\right)\nonumber\\
&=& \int dm_b^2\ \Delta_{b}(i\omega_n,{\mathbf {k}};|q_bB|),
\label{using}
\end{eqnarray}
we obtain
\begin{eqnarray}
V_{b}^{(1;B)}\!\!&=&\!\!\frac{T}{2}\sum_n\int dm_b^2\int \frac{d^3k}{(2\pi)^{3}}\int_0^\infty\!\! \frac{ds}{\cosh |q_bB|s}\nonumber \\
&\times& \exp\left\{
-s\left(\omega_n^2+k_3^2+k_\perp^2\frac{\tanh|q_bB|s}{|q_bB|s}+m_b^2\right)
\right\}.\nonumber\\
\label{effoneloopBexpl}
\end{eqnarray}
Performing the integration over $k_\perp$, introducing the sum over Landau levels, integrating over $s$, performing the sum over Matsubara frequencies and the integration over $m_b^2$, in that order, we get
\begin{eqnarray}
V_{b}^{(1;B)}=\frac{|q_bB|}{4\pi}\sum_l\int_{-\infty}^{\infty}\frac{dk_3}{2\pi}\left[\omega_l + 2T\ln\left(1-e^{-\omega_l/T}\right)\right],\nonumber \\
\label{1loopVBafter}
\end{eqnarray}
where
\begin{eqnarray}
\omega_l=\sqrt{k_3^2+m_b^2+(2l+1)|q_bB|}.
\label{fequencies}
\end{eqnarray}
Notice that Eq.~(\ref{1loopVBafter}) splits into vacuum and matter contributions, namely
\begin{eqnarray}
V_{b}^{(1;B\ vac)}=\frac{|q_bB|}{4\pi}\sum_l\int_{-\infty}^{\infty}\frac{dk_3}{2\pi}\omega_l,
\label{1loopbosvac}
\end{eqnarray}
and
\begin{eqnarray}
V_{b}^{(1;B\ matt)}=\frac{2|q_bB|}{4\pi}T\sum_l\int_{-\infty}^{\infty}\frac{dk_3}{2\pi}\ln\left(1-e^{-\omega_l/T}\right). \nonumber \\
\label{1loopbosmat}
\end{eqnarray}

We now specialize to implementing the calculation having in mind the conditions after a relativistic heavy-ion collision. Recall that in this environment, although the magnetic field intensity can be initially very large, it decreases fast such that, for the time the plasma reaches thermal equilibrium, it becomes weak. Under such conditions, it seems plausible to compute the effective potential, in the weak field limit. For this purpose let us write Eq.~(\ref{1loopVBafter}) as
\begin{eqnarray}
V_{b}^{(1;B\ vac)}=\frac{S_b}{2},
\label{EM1}
\end{eqnarray}
where
\begin{eqnarray}
S_b&\equiv&\sum_l \frac{h}{4\pi}f_l\nonumber\\
f_l&\equiv&\int_{-\infty}^{\infty}\frac{dk_3}{2\pi}\omega_l,
\label{EM2}
\end{eqnarray}
with $h=2|q_bB|$.
The sum can be performed resorting to the Euler-Mclaurin formula, writing
\begin{eqnarray}
\!\!\!\! h\left[\frac{f_0}{2}+f_1+f_2+\ldots +\frac{f_N}{2}\right]&=&\int_0^{Nh}dxf(x)\nonumber \\
&+& \frac{B_2}{2!}h^2(f'_N-f'_0),
\label{EM3}
\end{eqnarray}
where $B_2=1/6$ is the second Bernoulli number and we have kept terms only up to ${\mathcal{O}}(q_bB)^2$. The limit $N\to\infty$ is to be understood. Notice that on the right-hand side of Eq.~(\ref{EM3}) we made the replacement $2|q_bB|(l+1/2)\to x$, which means that, if we think of the series of $f_l$ as being represented by a histogram, we are effectively performing the sum evaluating the function at the middle point of each bar in the histogram. This means that on the left-hand side of Eq.~(\ref{EM3}),  the first and last terms in the sum are weighed with $h/2$. For small $|q_bB|$, $x$ can be thought of as being continuous, and the derivative is taken with respect to this variable. The meaning of $x$ can in turn be made more appealing by writing
\begin{eqnarray}
x&=&k_\perp^2,\nonumber\\
dx=dk_\perp^2&=&2k_\perp dk_\perp, 
\label{rewrite}
\end{eqnarray}
therefore
\begin{eqnarray}
\int_0^{Nh}dxf(x)\to 2(2\pi)\int\frac{d^3k}{(2\pi)^3}\sqrt{k^2+m_b^2},
\label{intf}
\end{eqnarray}
and
\begin{eqnarray}
f'\equiv\frac{df}{dk_\perp^2}&=&\frac{1}{2}\int_{-\infty}^\infty\frac{dk_3}{2\pi}\frac{1}{\sqrt{k^2+m_b^2}}\nonumber\\
(f'_\infty -f'_0)&=&-\frac{1}{2}\int_{-\infty}^\infty\frac{dk_3}{2\pi}\frac{1}{\sqrt{k_3^2+m_b^2}}.
\label{secondterm}
\end{eqnarray}
Hence, bringing all together we get
\begin{eqnarray}
V_{b}^{(1;B\ vac)}&=&\frac{1}{2}\int\frac{d^3k}{(2\pi)^3}\sqrt{k^2+m_b^2}\nonumber \\
&-& \frac{|q_bB|^2}{48\pi}\int_{-\infty}^\infty\frac{dk_3}{2\pi}\frac{1}{\sqrt{k_3^2+m_b^2}}.
\label{together}
\end{eqnarray}
In order to find an explicit expression, we use dimensional regularization, introducing the ultraviolet renormalization scale $\tilde{\mu}$ in the $\overline{\mbox{MS}}$ scheme, to get
\begin{eqnarray}
V_{b}^{(1;B\ vac)}&=&\frac{m_b^4}{64\pi^2}\left[\ln\left(\frac{m_b^2}{\tilde{\mu}^2}\right)-\frac{1}{\epsilon}-\frac{3}{2}
\right]\nonumber \\
&+&\frac{|q_bB|^2}{96\pi^2}\left[\ln\left(\frac{m_b^2}{\tilde{\mu}^2}\right)-\frac{1}{\epsilon}\right].
\label{dimreg}
\end{eqnarray}
Notice that the divergent piece in the first line of Eq.~(\ref{dimreg}) is harmless as it is $T$ and $B$-independent and can be safely ignored. On the other hand, the divergent term in the second line of Eq.~(\ref{dimreg}) is potentially dangerous since it is $B$-dependent. However, recall that the vacuum mass divergence is cured precisely by the inclusion of a mass counterterm $\delta m^2 \sim m^2/\epsilon$. Therefore, the addition of this counterterm in the second line of Eq.~(\ref{dimreg}) effectively induces the substitution
\begin{eqnarray}
\ln\left[\frac{m_b^2}{\tilde{\mu}^2}\right]&\to&\ln\left[\frac{m_b^2}{\tilde{\mu}^2}\left(1+\frac{1}{\epsilon}\right)\right]\nonumber\\
&\sim&\ln\left[\frac{m_b^2}{\tilde{\mu}^2}\right]+\frac{1}{\epsilon}.
\label{induces}
\end{eqnarray}
Therefore, the renormalized vacuum contribution is written as
\begin{eqnarray}
V_{b}^{(1;B\ ren)}&=&\frac{m_b^4}{64\pi^2}\left[\ln\left(\frac{m_b^2}{\tilde{\mu}^2}\right)-\frac{3}{2}
\right]\nonumber \\
&+&\frac{|q_bB|^2}{96\pi^2}\left[\ln\left(\frac{m_b^2}{\tilde{\mu}^2}\right)\right].
\label{renvac}
\end{eqnarray}

We now look at the matter contribution to the one-loop effective potential, Eq.~(\ref{1loopbosmat}), which we write as
\begin{eqnarray}
V_{b}^{(1;B\ matt)}=TS_b^{matt},
\label{mattB}
\end{eqnarray}
where
\begin{eqnarray}
\label{SmattB}
S_b^{matt}&=&\sum_l\frac{h}{4\pi}g_l,\nonumber\\
g_l&\equiv&\int_{-\infty}^\infty\frac{dk_3}{2\pi}\ln\left(1-e^{-\omega_l/T}\right).
\end{eqnarray}
Once again, working in the weak field limit we can write
\begin{eqnarray}
\!\!\!\! h\left[\frac{g_0}{2}+g_1+g_2+\ldots +\frac{g_N}{2}\right]&=&\int_0^{Nh}dxg(x) \nonumber \\
&+& \frac{B_2}{2!}h^2(g'_N-g'_0).
\label{EMmatt1}
\end{eqnarray}
In the limit $N\to\infty$
\begin{eqnarray}
\int_0^{Nh}dx\ g(x)\to 2(2\pi)\int\frac{d^3k}{(2\pi)^3}\ln\left(1-e^{-\sqrt{k^2+m_b^2}/T}\right),\nonumber \\
\label{intg}
\end{eqnarray}
and
\begin{eqnarray}
g'\equiv\frac{dg}{dk_\perp^2}&=&\frac{1}{2T}\int_{-\infty}^\infty\frac{dk_3}{2\pi}\frac{1}{\sqrt{k^2+m_b^2}}\left(\frac{1}{e^{\sqrt{k^2+m_b^2}/T}-1}\right),\nonumber\\
(g'_\infty -g'_0)&=&-\frac{1}{2T}\int_{-\infty}^\infty\frac{dk_3}{2\pi}\frac{1}{\sqrt{k_3^2+m_b^2}}\left(\frac{1}{e^{\sqrt{k_3^2+m_b^2}/T}-1}\right). \nonumber \\
\label{secondtermg}
\end{eqnarray}
Thus
\begin{eqnarray}
V_{b}^{(1;B\ matt)}&=&T\int\frac{d^3k}{(2\pi)^3}\ln\left(1-e^{-\sqrt{k^2+m_b^2}/T}\right)\nonumber\\
&-&\frac{|q_bB|^2}{24\pi}\int_{-\infty}^\infty\frac{dk_3}{2\pi}\frac{1}{\sqrt{k_3^2+m_b^2}}\nonumber\\
&\times&\left(\frac{1}{e^{\sqrt{k_3^2+m_b^2}/T}-1}\right).
\label{togetherg}
\end{eqnarray}

We now proceed to provide an  approximate expression for Eq.~(\ref{togetherg}) in the large $T$-limit. The first term is given by~\cite{Dolan:1973qd}

\begin{widetext}
\begin{eqnarray}
T\int\frac{d^3k}{(2\pi)^3}\ln\left(1-e^{-\sqrt{k^2+m_b^2}/T}\right)\simeq-\frac{T^4\pi^2}{90}
+\frac{T^2m_b^2}{24}-\frac{Tm_b^3}{12\pi}
-\frac{m_b^4}{64\pi^2}\Bigg[\ln\left(\frac{m_b^2}{(4\pi T)^2}\right)+2\gamma_E-\frac{3}{2}\Bigg].
\label{largeTDol}
\end{eqnarray}
For the second term we get~\cite{Kapusta:1989tk}
\begin{eqnarray}
-\frac{|q_bB|^2}{24\pi}\int_{-\infty}^\infty\frac{dk_3}{2\pi}\frac{1}{\sqrt{k_3^2+m_b^2}}\left(\frac{1}{e^{\sqrt{k_3^2+m_b^2}/T}-1}\right)&\simeq& -\frac{|q_bB|^2}{24\pi^2}\left[\frac{T\pi}{2m_b}+\frac{1}{4}\ln\left(\frac{m_b^2}{(4\pi T)^2}\right)+\frac{1}{2}\gamma_E\right.\nonumber\\
&-&\left.\frac{1}{4}\zeta(3)\left(\frac{m_b^2}{(2\pi T)^2}\right)+\frac{3}{16}\zeta(5)\left(\frac{m_b^4}{(2\pi T)^4}\right)\right],\nonumber\\
\label{largeTKap}
\end{eqnarray}
\end{widetext}
where $\gamma_E$ and $\zeta$ are the Euler-Mascheroni constant and Riemann Zeta function, respectively. Notice that the arguments of the logarithms in Eqs.~(\ref{renvac}),~(\ref{largeTDol}) and~(\ref{renvac}) are potentially dangerous when $m_b^2$ becomes zero or even negative. However, when adding up these equations to express the one-loop boson contribution to the effective potential, these terms combine in such a way that the argument of the logarithms does not contain $m_b^2$ anymore. We thus obtain
\begin{widetext}
\begin{eqnarray}
V_{b}^{(1;B\ ren)}+V_{b}^{(1;B\ matt)}&=&-\frac{T^4\pi^2}{90}+\frac{T^2m_b^2}{24}-\frac{Tm_b^3}{12\pi}-\frac{m_b^4}{64\pi^2}\left[\ln\left(\frac{\tilde{\mu}^2}{(4\pi T)^2}\right)+2\gamma_E\right]\nonumber\\
&-&\frac{|q_bB|^2}{24\pi^2}\left[\frac{T\pi}{2m_b}+\frac{1}{4}\ln\left(\frac{\tilde{\mu}^2}{(4\pi T)^2}\right)+\frac{1}{2}\gamma_E-\frac{1}{4}\zeta(3)\left(\frac{m_b^2}{(2\pi T)^2}\right)+\frac{3}{16}\zeta(5)\left(\frac{m_b^4}{(2\pi T)^4}\right)\right].
\label{V1loopcomplete}
\end{eqnarray}
\end{widetext}

Notice however that Eq.~(\ref{V1loopcomplete}) contains odd powers of $m_b=\sqrt{m_b^2}$. When the boson mass squared is negative, these terms become imaginary. This instability is non-physical and, as we proceed to show, it can be cured by considering the contribution from the {\it ring diagrams}.

\subsection{Ring contribution}\label{secII-2}

Recall that from Eq.~(\ref{effring-2}), the ring contribution to the effective potential can be written as
\begin{eqnarray}
V_{b}^{(ring)}&=&
\frac{T}{2}\sum_n\int \frac{d^3k}{(2\pi)^{3}}\left(\ln\left[\Delta_{b}^{-1}(i\omega_n,{\mathbf {k}})+\Pi_b(i\omega_n,{\mathbf {k}})\right]\right. \nonumber\\
&-&\left. \ln\left[\Delta^{-1}_{b}(i\omega_n,{\mathbf {k}})
\right]\right).
\label{effring2}
\end{eqnarray}
At high temperature, the leading ring contribution comes from the $n=0$ Matsubara mode, so we write
\begin{eqnarray}
V_{b}^{(ring)}&=&
\frac{T}{2}\int \frac{d^3k}{(2\pi)^{3}}\left(\ln\left[\Delta_{b}^{-1}(i\omega_0,{\mathbf {k}})+\Pi_b\right]\right. \nonumber\\
&-&\left.\ln\left[\Delta^{-1}_{b}(i\omega_0,{\mathbf {k}})
\right]\right),
\label{effring3}
\end{eqnarray}
where we have also approximated $\Pi_b$ by a momentum independent quantity. This choice will be justified later on when we discuss the computation of the self-energy. Equation~(\ref{effring3}) contains two terms. The first one is obtained from the second by the replacement $m_b^2\to m_b^2 + \Pi_b$. Therefore we concentrate on the computation of the second term
\begin{widetext}
\begin{eqnarray}
\frac{T}{2}\int\frac{d^3k}{(2\pi)^3}\ln\left[\Delta^{-1}_b(i\omega_0,{\mathbf{k}})\right]&=&\frac{T}{2}\int\ dm_b^2\int\frac{d^3k}{(2\pi)^3} \frac{d}{dm_b^2}\ln\left[\Delta^{-1}_b(i\omega_0,{\mathbf{k}})\right]\nonumber\\
&=&\frac{T}{2}\int\ dm_b^2\int\frac{d^3k}{(2\pi)^3}\Delta_b(i\omega_0,{\mathbf{k}})\nonumber\\
&=&\frac{T}{2}\int\ dm_b^2\int\frac{d^3k}{(2\pi)^3}\int_0^\infty\!\!\!\! \frac{ds}{\cosh |q_bB|s}e^{-s\left(k_3^2+k_\perp^2\frac{\tanh|q_bB|s}{|q_bB|s}+m_b^2\right)}.\nonumber\\
\label{ringsec}
\end{eqnarray}
Performing the integration over $k_\perp$ we obtain
\begin{eqnarray}
\frac{T}{2}\int\frac{d^3k}{(2\pi)^3}\ln\left[\Delta^{-1}_b(i\omega_0,{\mathbf{k}})\right]=T\left(\frac{|q_bB|}{8\pi}\right)\int dm_b^2\int_{-\infty}^{\infty}\frac{dk_3}{2\pi}\int_0^\infty\frac{ds}{\sinh |q_bB|s}e^{-s\left(k_3^2+m_b^2\right)}.\nonumber\\
\label{overkperp}
\end{eqnarray}
Introducing the series representation
\begin{eqnarray}
\!\!\!\!\!\!\!\!\frac{1}{\sinh(x)}=2\sum_{l=0}^{\infty}e^{-(2l+1)x},
\label{seriesrep}
\end{eqnarray}
we get
\begin{eqnarray}
\frac{T}{2}\int\frac{d^3k}{(2\pi)^3}\ln\left[\Delta^{-1}_b(i\omega_0,{\mathbf{k}})\right]=T\left(\frac{|q_bB|}{4\pi}\right)\int dm_b^2\int_{-\infty}^{\infty}\frac{dk_3}{2\pi}\sum_{l=0}^\infty\int_0^\infty \!\!\!ds\ e^{-s\left(k_3^2+m_b^2+(2l+1)|q_bB|\right)}.
\label{afterseries}
\end{eqnarray}
Carrying out the integrations over $s$ and $k_3$, we have
\begin{eqnarray}
\frac{T}{2}\int\frac{d^3k}{(2\pi)^3}\ln\left[\Delta^{-1}_b(i\omega_0,{\mathbf{k}})\right]=T\left(\frac{|q_bB|}{8\pi}\right)\sum_{l=0}^\infty\int \frac{dm_b^2}{\sqrt{m_b^2+(2l+1)|q_bB|}}
=T\left(\frac{2|q_bB|}{8\pi}\right)\sum_{l=0}^\infty\sqrt{m_b^2+(2l+1)|q_bB|}.\nonumber\\
\label{aftersk3}
\end{eqnarray}
For the weak field case, we can once again use the Euler-Maclaurin formula to write
\begin{eqnarray}
\frac{T}{2}\int\frac{d^3k}{(2\pi)^3}\ln\left[\Delta^{-1}_b(i\omega_0,{\mathbf{k}})\right]&=&\left(\frac{T}{8\pi}\right)\left\{\int_0^{\Lambda^2}dx\sqrt{m_b^2+x}-\frac{|q_bB|^2}{6m_b}\right\}=\left(\frac{T}{8\pi}\right)\left\{\frac{2}{3}\left(m_b^2+\Lambda^2\right)^{3/2}-\frac{2}{3}m_b^2-\frac{|q_bB|^2}{6m_b}\right\}\nonumber\\
&\simeq&\left(\frac{T}{12\pi}\right)\left\{\Lambda^3 - m_b^3 - \frac{|q_bB|^2}{4m_b}\right\},
\label{EM-ring}
\end{eqnarray}
\end{widetext}
where, since the integral on the right-hand side diverges, the upper limit of integration has been set as the large but finite quantity $\Lambda^2$. The dependence of the result on $\Lambda$ will drop out wen including the $\Pi_b$-dependent term in Eq.~(\ref{effring2}). Using a similar procedure to treat the first term of Eq.~(\ref{effring3}), we obtain
\begin{eqnarray}
V_b^{(ring)}&=&\left(\frac{T}{12\pi}\right)\left\{m_b^3 - (m_b^2+\Pi_b)^{3/2}\right. \nonumber\\&+&\left.\frac{|q_bB|^2}{4m_b}
- \frac{|q_bB|^2}{4(m_b^2+\Pi_b)^{1/2}}\right\},
\label{ringtot}
\end{eqnarray}
where, as promised, the dependence on $\Lambda$ has dropped out. Therefore, adding this contribution to Eq.~(\ref{V1loopcomplete}), we get
\begin{eqnarray}
V_b^{(eff)}&\equiv& V_{b}^{(1;B\ ren)}+V_{b}^{(1;B\ matt)}+V_b^{(ring)}=\nonumber \\
&-&\frac{T^4\pi^2}{90}+\frac{T^2m_b^2}{24}-\frac{T(m_b^2+\Pi_b)^{3/2}}{12\pi}\nonumber\\
&-&\frac{m_b^4}{64\pi^2}\left[\ln\left(\frac{\tilde{\mu}^2}{(4\pi T)^2}\right)+2\gamma_E\right]\nonumber\\
&-&\frac{|q_bB|^2}{24\pi^2}\left[\frac{T\pi}{2(m_b^2+\Pi_b)^{1/2}}+\frac{1}{4}\ln\left(\frac{\tilde{\mu}^2}{(4\pi T)^2}\right)\right.\nonumber\\
&+&\left.\frac{1}{2}\gamma_E-\frac{1}{4}\zeta(3)\left(\frac{m_b^2}{(2\pi T)^2}\right)+\frac{3}{16}\zeta(5)\left(\frac{m_b^4}{(2\pi T)^4}\right)\right]\!.\nonumber\\
\label{V1loopcomplete2}
\end{eqnarray}

We thus confirm that the inclusion of the ring diagrams produces an effective potential which is free of infrared instabilities. In order to use this expression to study the phase diagram, it is necessary to compute the boson self-energy $\Pi_b$. Before we proceed to this calculation, we first compute the contribution from fermions to the effective potential.

\subsection{One-loop fermion contribution}\label{secII-3}

We start from the expression for the one-loop effective potential for one fermion species. Working first in Minkowsky space, this expression is given by
\begin{eqnarray}
V_f^{(1;B)}&=&i\ln{\mbox{Det}}(iS_f^{-1})\nonumber\\
&=&i{\mbox{Tr}}\ln(iS_f^{-1})\nonumber\\
&=&i{\mbox{Tr}}\ln(\slashed{\Pi}-m_f),
\label{V1loopfer}
\end{eqnarray}
where $S_f$ is the fermion propagator in the presence of a magnetic field, $\Pi_\mu=p_\mu-q_fA_\mu$ is the kinematical momentum and $q_f$, $m_f$ are the fermion electric charge and mass, respectively. In order to find a working expression, notice that
\begin{eqnarray}
{\mbox{Det}}(\slashed{\Pi}+m_f){\mbox{Det}}(\slashed{\Pi}-m_f)&=&{\mbox{Det}}\left[(\slashed{\Pi}+m_f)(\slashed{\Pi}-m_f)\right]\nonumber\\
&=&{\mbox{Det}}
\left[\slashed{\Pi}^2-m_f^2\right],
\label{noticefer}
\end{eqnarray}
then
\begin{eqnarray}
\ln{\mbox{Det}}\left[\slashed{\Pi}^2-m_f^2\right]=\ln{\mbox{Det}}(\slashed{\Pi}+m_f)+\ln{\mbox{Det}}(\slashed{\Pi}-m_f).\nonumber\\
\label{further1}
\end{eqnarray}
Also, since in four space-time dimensions
\begin{eqnarray}
{\mbox{Det}}(\slashed{\Pi}+m_f)&=&{\mbox{Det}}\left[\gamma_5^2(\slashed{\Pi}+m_f)\right]\nonumber\\
&=&{\mbox{Det}}\left[\gamma_5(-\slashed{\Pi}+m_f)\gamma_5\right]\nonumber\\
&=&{\mbox{Det}}(\slashed{\Pi}-m_f),
\label{further2}
\end{eqnarray}
we have
\begin{eqnarray}
\ln{\mbox{Det}}(\slashed{\Pi}+m_f)+\ln{\mbox{Det}}(\slashed{\Pi}-m_f)=2\ln{\mbox{Det}}(\slashed{\Pi}-m_f). \nonumber\\
\label{further3}
\end{eqnarray}
Therefore, using Eq.~(\ref{further1}), we have
\begin{eqnarray}
\ln{\mbox{Det}}(\slashed{\Pi}-m_f)=\frac{1}{2}\ln{\mbox{Det}}(\slashed{\Pi}^2-m_f^2),
\label{further4}
\end{eqnarray}
and all in all, we get
\begin{eqnarray}
V_f^{(1;B)}=\frac{i}{2}{\mbox{Tr}}\ln(\slashed{\Pi}^2-m_f^2).
\label{further5}
\end{eqnarray}
Recall that 
\begin{eqnarray}
\slashed{\Pi}^2=\Pi^2-q_fB\Sigma_3,
\label{further6}
\end{eqnarray}
where $\Sigma_3$ is the spin operator along the $\hat{z}$-axis. We thus see that after accounting for the trace, the expression for the fermion contribution to the one-loop effective potential is equivalent to minus the contribution from four scalars, two of them carrying a label representing the spin projection parallel and the other two antiparallel to the magnetic field. The spin projection can be more easily implemented introducing a sum over an index $\sigma$ that takes on the values $\pm 1$. Since both components are involved, it is enough to write $q_fB\Sigma_3\to |q_fB|\sigma$. Therefore, the expression for the fermion contribution to the one-loop effective potential at finite temperature is written as
\begin{eqnarray}
\!\!\!\!\!\!V_{f}^{(1;B)}\!&=&\!-2\sum_{\sigma=\pm 1}\frac{T}{2}\sum_n\int dm_f^2\int \frac{d^3k}{(2\pi)^{3}}
\int_0^\infty\!\!\!\! \frac{ds}{\cosh |q_bB|s}\nonumber\\
&\times&e^{
-s\left[(\tilde{\omega}_n+i\mu)^2+k_3^2+k_\perp^2\frac{\tanh|q_bB|s}{|q_bB|s}+m_f^2+\sigma |q_fB|\right]},\nonumber\\
\label{Vfer1loop}
\end{eqnarray}
where $\tilde{\omega}_n=(2n+1)\pi T$ is a fermion Matsubara frequency and $\mu$ is the quark chemical potential. The factor of 2 and the sum over $\sigma$ take care of the four fermion degrees of freedom.

Performing the integration over $k_\perp$, introducing the sum over Landau levels, integrating over $s$, performing the sum over Matsubara frequencies and the integration over $m_f^2$, in that order, we get
\begin{eqnarray}
V_{f}^{(1;B)}&=&-\frac{2|q_fB|}{4\pi}\sum_{l,\sigma}\int_{-\infty}^{\infty}\frac{dk_3}{2\pi}\left[\omega_l \right.\nonumber\\
&+& T\ln\left(1+e^{-(\omega_l-\mu)/T}\right)\nonumber\\
&+&\left.T\ln\left(1+e^{-(\omega_l+\mu)/T}\right)\right],
\label{1loopVBfafter}
\end{eqnarray}
where
\begin{eqnarray}
\omega_l=\sqrt{k_3^2+m_f^2+(2l+1+\sigma)|q_fB|}.
\label{fequenciesf}
\end{eqnarray}

Once again in Eq.~(\ref{fequenciesf}) we distinguish two kinds of terms, vacuum and matter contributions, namely
\begin{eqnarray}
V_{f}^{(1;B\ vac)}=-\frac{2|q_fB|}{4\pi}\sum_{l,\sigma}\int_{-\infty}^{\infty}\frac{dk_3}{2\pi}\omega_l,
\label{fervac}
\end{eqnarray}
and
\begin{eqnarray}
V_{f}^{(1;B\ matt)}&=&-\frac{2|q_fB|}{4\pi}T\sum_{l,\sigma}\int_{-\infty}^{\infty}\frac{dk_3}{2\pi}\nonumber\\ &\times&\left[\ln\left(1+e^{-(\omega_l-\mu)/T}\right) \right. \nonumber\\
&+&\left. \ln\left(1+e^{-(\omega_l+\mu)/T}\right)\right].
\label{fermat}
\end{eqnarray}

We now proceed to compute each of these contributions separately. We start with the vacuum contribution. Writing the sum over $\sigma$ explicitly, we get
\begin{eqnarray}
\!\!\!\!\!V_{f}^{(1;B\ vac)}\!&=&\!-\frac{2|q_fB|}{4\pi}\sum_{l=0}^{\infty}\int_{-\infty}^{\infty}\frac{dk_3}{2\pi}\nonumber\\
&\times&\left[\sqrt{k_3^2+m_f^2+2(l+1)|q_fB|}\right. \nonumber\\
&+&\left. \sqrt{k_3^2+m_f^2+2l|q_fB|}\right].
\label{vacexpl}
\end{eqnarray}
We now separate the term with $l=0$ writing
\begin{eqnarray}
V_{f}^{(1;B\ vac)}&=&-\frac{2|q_fB|}{4\pi}\int_{-\infty}^{\infty}\frac{dk_3}{2\pi}\left[\sqrt{k_3^2+m_f^2}\right. \nonumber\\
&+&2\left. \sum_{l=1}^\infty\sqrt{k_3^2+m_f^2+2l|q_fB|}\right].
\label{vacexpl2}
\end{eqnarray}
Let us first compute the term with the sum. Define
\begin{eqnarray}
S_f&\equiv& \sum_{l=1}^\infty\frac{h}{4\pi}f_l\nonumber\\
f_l&\equiv&2\int_{-\infty}^\infty \frac{dk_3}{2\pi}\sqrt{k_3^2+m_f^2+2l|q_fB|},
\label{sumferI}
\end{eqnarray}
where $h=2|q_fB|$. Notice that, were we to represent the series of $f_l$ by a histogram, the function would then be evaluated at the edges of each bar. Thus, the expression for the Euler-Maclaurin formula to use is
\begin{eqnarray}
h\left[f_1+f_2+\ldots + \frac{f_N}{2}\right]&=&\int_0^{Nh}dxf(x)-h\frac{f(0)}{2}\nonumber\\
&+&\frac{B_2}{2!}h^2(f'_N-f'_0),
\label{correctEM}
\end{eqnarray}
where again, $B_2=1/6$ and we have kept terms only up to ${\mathcal{O}}(q_fB)^2$. Notice that we have maintained the last term on the left-hand side as $f_N/2$ since in the limit when $N\to\infty$ this does not make a difference and the divergence is anyway taken care of by using dimensional regularization for the expression on the right-hand side. Also, in Eq.~(\ref{correctEM}), the contribution from the term $-f(0)/2$ cancels the first term on the right-hand side of Eq.~(\ref{vacexpl2}).

We now make the change of variable in Eq.~(\ref{rewrite}) to write
\begin{eqnarray}
\int_0^{Nh}dxf(x)\to 2(4\pi)\int\frac{d^3k}{(2\pi)^3}\sqrt{k^2+m_f^2}
\label{intfer}
\end{eqnarray}
and
\begin{eqnarray}
f'\equiv\frac{df}{dk_\perp^2}&=&\int_{-\infty}^\infty\frac{dk_3}{2\pi}\frac{1}{\sqrt{k^2+m_f^2}}\nonumber\\
(f'_\infty -f'_0)&=&-\int_{-\infty}^\infty\frac{dk_3}{2\pi}\frac{1}{\sqrt{k_3^2+m_f^2}}.
\label{secondtermfer}
\end{eqnarray}
Writing these ingredients all together, we have
\begin{eqnarray}
V_{f}^{(1;B\ vac)}&=&-2\int\frac{d^3k}{(2\pi)^3}\sqrt{k^2+m_f^2}\nonumber\\
&-& \frac{|q_fB|^2}{12\pi}\int_{-\infty}^\infty\frac{dk_3}{2\pi}\frac{1}{\sqrt{k_3^2+m_f^2}}.
\label{alltogfer}
\end{eqnarray}
Therefore, working in the $\overline{\mbox{MS}}$ scheme and after accounting for the fermion mass renormalization, we get
\begin{eqnarray}
V_{f}^{(1;B\ ren)}&=&\frac{m_f^4}{16\pi^2}\left[\ln\left(\frac{\tilde{\mu}^2}{m_f^2}\right)+\frac{3}{2}
\right]\nonumber\\
&+&\frac{|q_fB|^2}{24\pi^2}\left[\ln\left(\frac{\tilde{\mu}^2}{m_f^2}\right)\right].
\label{ferrenvac}
\end{eqnarray}

We now turn to the calculation of the matter contribution, Eq.~(\ref{fermat}), which, after accounting for the sum over $\sigma$ and separating the term of the sum with $l=0$ can be written as
\begin{eqnarray}
V_{f}^{(1;B\ matt)}&=&-\frac{2|q_fB|}{4\pi}T\int_{-\infty}^{\infty}\frac{dk_3}{2\pi}\nonumber\\
&\times&\Big\{
\left[\ln\left(1+e^{-\left(\sqrt{k_3^2+m_f^2}-\mu\right)/T}\right) \right.\nonumber\\
&+&\left. \ln\left(1+e^{-\left(\sqrt{k_3^2+m_f^2}+\mu\right)/T}\right)\right]\nonumber\\
&+&\sum_{l=1}^\infty \left[\ln\left(1+e^{-\left(\sqrt{k_3^2+m_f^2+2l|q_fB|}-\mu\right)/T}\right) \right. \nonumber\\
&+&\left. \ln\left(1+e^{-\left(\sqrt{k_3^2+m_f^2+2l|q_fB|}+\mu\right)/T}\right)\right]\Big\}.\nonumber\\
\label{fermatafter0}
\end{eqnarray}
In order to employ the Euler-Maclaurin formula up to $\mathcal{O}(q_fB)^2$, we write
\begin{eqnarray}
h\left[g_1+\ldots + g_N\right]&=&\int_0^{Nh} dx g(x) + h\frac{g(Nh)-g(0)}{2}\nonumber\\
&+& \frac{B_2}{2!}h^2(g'_N-g'_0),
\label{EMfer1}
\end{eqnarray}
and identify 
\begin{eqnarray}
\!\!\!\!\!\!\!g_l&=&2\int_{-\infty}^\infty\frac{dk_3}{2\pi}\left[\ln\left(1+e^{-\left(\sqrt{k_3^2+m_f^2+2l|q_fB|}-\mu\right)/T}\right)\right. \nonumber\\
&+&\left. \ln\left(1+e^{-\left(\sqrt{k_3^2+m_f^2+2l|q_fB|}+\mu\right)/T}\right)\right],
\label{EMfer2}
\end{eqnarray}
with $h=2|q_fB|$. Notice that
\begin{eqnarray}
g_\infty&=&0\nonumber\\
g_0&=&2\int_{-\infty}^{\infty}\frac{dk_3}{2\pi}
\left[\ln\left(1+e^{-\left(\sqrt{k_3^2+m_f^2}-\mu\right)/T}\right)\right. \nonumber\\
&+&\left. \ln\left(1+e^{-\left(\sqrt{k_3^2+m_f^2}+\mu\right)/T}\right)\right],
\label{EMfer3}
\end{eqnarray}
and we obtain
\begin{eqnarray}
V_{f}^{(1;B\ matt)}&=&-2\ T\int\frac{d^3k}{(2\pi)^3}\left[\ln\left(1+e^{-\left(\sqrt{k^2+m_f^2}-\mu\right)/T}\right) \right. \nonumber\\
&+&\left. \ln\left(1+e^{-\left(\sqrt{k^2+m_f^2}+\mu\right)/T}\right)\right]\nonumber\\
&-&\frac{|q_fB|^2}{12\pi}\int_{-\infty}^{\infty}\frac{dk_3}{2\pi}\frac{1}{\sqrt{k_3^2+m_f^2}}\nonumber\\
&\times&
\left[\frac{1}{e^{\left(\sqrt{k_3^2+m_f^2}-\mu\right)/T}+1}\right.\nonumber\\
&+&\left. \frac{1}{e^{\left(\sqrt{k_3^2+m_f^2}+\mu\right)/T}+1}\right].\nonumber \\
\label{Vmatferfin}
\end{eqnarray}
We now look for a large-$T$ approximation for Eq.~(\ref{Vmatferfin}). Let us start working the second term. We write
\begin{eqnarray}
x&=&\frac{k_3}{T}\nonumber\\
y&=&\frac{m_f}{T}\nonumber\\
z&=&\frac{\mu}{T},
\label{secondfermat}
\end{eqnarray}
and use that

\begin{widetext}
\begin{eqnarray}
\frac{1}{\sqrt{x^2+y^2}}\left\{\frac{1}{e^{\sqrt{x^2+y^2}-z}}+\frac{1}{e^{\sqrt{x^2+y^2}+z}}\right\}&=&\frac{1}{\sqrt{x^2+y^2}}-2\sum_{n=-\infty}^\infty\frac{1}{\left[\left(2n+1\right)\pi + iz\right]^2+x^2+y^2}\nonumber\\
&=&\frac{1}{\sqrt{x^2+y^2}}-2\sum_{n=-\infty}^\infty\!\frac{1}{\left(\left[\left(2n+1\right)\pi + iz\right]^2+y^2\right)\left(1+\frac{x^2}{\left[\left(2n+1\right)\pi + iz\right]^2+y^2}\right)}.\nonumber\\
\label{identi}
\end{eqnarray}
Thus, the integral in the second term of Eq.~(\ref{Vmatferfin}) can be expressed in terms of two terms given by
\begin{eqnarray}
I_1&=&\int_0^\infty\frac{x^\epsilon dx}{\sqrt{x^2+y ^2}}\nonumber\\
I_2&=&-2\sum_{n=-\infty}^\infty\int_0^\infty\frac{x^\epsilon dx}{\left(\left[\left(2n+1\right)\pi + iz\right]^2+y^2\right)\left(1+\frac{x^2}{\left[\left(2n+1\right)\pi + iz\right]^2+y^2}\right)},
\label{seconprop2}
\end{eqnarray}
with
\begin{eqnarray}
\int_{-\infty}^{\infty}\frac{dk_3}{2\pi}\frac{1}{\sqrt{k_3^2+m_f^2}}
\left[\frac{1}{e^{\left(\sqrt{k_3^2+m_f^2}-\mu\right)/T}}+\frac{1}{e^{\left(\sqrt{k_3^2+m_f^2}+\mu\right)/T}}\right]=\frac{1}{\pi}(I_1+I_2).
\label{sumI1I2}
\end{eqnarray}
\end{widetext}
Notice that, in anticipation to treating the divergence of each term, we have introduced the regulating factor $x^\epsilon$. As we proceed to show, this divergence cancels when adding both terms. First, notice that
\begin{eqnarray}
I_1&=&\frac{y^{-\epsilon}}{2\sqrt{\pi}}\Gamma\left[\frac{1}{2}-\frac{\epsilon}{2}\right]\nonumber\\
&=&\frac{1}{\epsilon} - \frac{\gamma_E}{2} - \frac{1}{2} \psi^0\left(\frac{1}{2}\right)-\ln{(y)}.
\label{I1fer}
\end{eqnarray}
For the computation of $I_2$, we define
\begin{eqnarray}
\omega_n^2=\left[(2n+1)\pi+iz\right]^2,
\label{defomegan}
\end{eqnarray}
and introduce the change of variable
\begin{eqnarray}
u=\frac{x}{\sqrt{\omega_n^2+y^2}},
\label{changevaru}
\end{eqnarray}
to write
\begin{eqnarray}
I_2=-2\sum_{n=-\infty}^{\infty}\frac{1}{\left(\omega^2+y^2\right)^{\frac{1+\epsilon}{2}}}\int_0^\infty
du\frac{u^{\epsilon}}{(1+u^2)}\label{towriteI2}.
\end{eqnarray}
The integral can be expressed as
\begin{eqnarray}
\int_0^\infty
du\frac{u^{\epsilon}}{(1+u^2)}\label{towriteI23}=\frac{\pi}{2}\sec\left(\frac{\pi\epsilon}{2}\right)=\frac{\pi}{2},
\label{usethatintu}
\end{eqnarray}
where we used that for $\epsilon\to 0$, the series of $\sec(\epsilon)$ starts at ${\mathcal{O}}(\epsilon)^2$ and thus its $\epsilon$ dependence can be discarded. Therefore
\begin{eqnarray}
I_2&=&-\pi\sum_{n=-\infty}^\infty\frac{1}{\left(\omega_n^2+y^2\right)^\frac{1+\epsilon}{2}}\nonumber\\
&=&-\pi\sum_{s=\pm 1}\sum_{n=0}^\infty\frac{1}{\left(\left[\left(2n+1\right)\pi+isz\right]^2+y^2\right)^\frac{1+\epsilon}{2}}.
\label{I2aft1}
\end{eqnarray}
To look for an expression to ${\mathcal{O}}(y^2)$, we expand the denominator of Eq.~(\ref{I2aft1}) to obtain

\begin{widetext}
\begin{eqnarray}
\frac{1}{\left(\left[\left(2n+1\right)\pi+isz\right]^2+y^2\right)^\frac{1+\epsilon}{2}}=\frac{1}{\left[\left(2n+1\right)\pi+isz\right]^\frac{1+\epsilon}{2}}-\frac{(1+\epsilon)}{2}\frac{y^2}{\left[\left(2n+1\right)\pi+isz\right]^{3+\epsilon}},\nonumber\\
\label{afterexp}
\end{eqnarray}
\end{widetext}
therefore
\begin{eqnarray}
I_2&=&-\pi\sum_{s=\pm 1}\left\{\frac{1}{(2\pi)^{1+\epsilon}}\zeta\left(1+\epsilon,\frac{1}{2}+\frac{isz}{2\pi}\right)\right. \nonumber\\
&-&\left.\frac{(1+\epsilon)}{2}\frac{y^2}{(2\pi)^{3+\epsilon}}\zeta\left(3+\epsilon,\frac{1}{2}+\frac{isz}{2\pi}\right)\right\},
\label{aftersumn}
\end{eqnarray}
where $\zeta$ is the Hurwitz Zeta function. Expanding for small $\epsilon$ we get
\begin{eqnarray}
I_2&=&-\frac{1}{\epsilon}+2\ln(2\pi)-\psi^0\left(\frac{1}{2}+\frac{iz}{2\pi}\right)- \psi^0\left(\frac{1}{2}-\frac{iz}{2\pi}\right)\nonumber\\
&+&\frac{y^2}{16\pi^2}\left[\zeta\left(3,\frac{1}{2}+\frac{iz}{2\pi}\right)+\zeta\left(3,\frac{1}{2}-\frac{iz}{2\pi}\right)\right],
\label{afterexpeps}
\end{eqnarray}
where $\psi^0$ is the digamma function. Adding up Eqs.~(\ref{I1fer}) and~(\ref{afterexpeps}) we obtain 
\begin{eqnarray}
&-&\frac{|q_fB|^2}{12\pi}\int_{-\infty}^{\infty}\frac{dk_3}{2\pi}\frac{1}{\sqrt{k_3^2+m_f^2}}\nonumber\\
&\times&\left[\frac{1}{e^{\left(\sqrt{k_3^2+m_f^2}-\mu\right)/T}}
+ \frac{1}{e^{\left(\sqrt{k_3^2+m_f^2}+\mu\right)/T}}\right]\nonumber\\
&=&-\frac{|q_fB|^2}{12\pi^2}\Bigg\{2\ln(2\pi)-\gamma_E-\frac{1}{2} \psi^0\left(\frac{1}{2}\right) \nonumber\\
&-&\frac{1}{2} \psi^0\left(\frac{1}{2}+\frac{i\mu}{2\pi T}\right)-\frac{1}{2} \psi^0\left(\frac{1}{2}-\frac{i\mu}{2\pi T}\right)\nonumber\\
&+&\frac{1}{2}\ln\left(\frac{4\pi^2T^2}{m_f^2}\right)+\frac{m_f^2}{16\pi^2T^2}\left[\zeta\left(3,\frac{1}{2}+\frac{i\mu}{2\pi T}\right) \right.  \nonumber\\
&+& \left. \zeta\left(3,\frac{1}{2}-\frac{i\mu}{2\pi T}\right)\right]\Bigg\}.
\label{casifinfer}
\end{eqnarray}
To evaluate the first term in Eq.~(\ref{Vmatferfin}) in the large-$T$ limit, we carry out a similar procedure. We first notice that in order to obtain an expansion of the integral up to $\mathcal{O}(m_f^4)$, we can take the second derivative of
\begin{eqnarray}
\!\!\!\!J(m_f/T)&\equiv&-2T\int\frac{d^3k}{(2\pi)^3}\left[\ln\left(1+e^{-\left(\sqrt{k^2+m_f^2}-\mu\right)/T}\right) \right. \nonumber\\
&+&\left. \ln\left(1+e^{-\left(\sqrt{k^2+m_f^2}+\mu\right)/T}\right)\right],
\label{firstlargeT}
\end{eqnarray}
with respect to $m_f^2$ and look for an approximation for $d^2J/(dm_f^2)^2$ up to the desired order. This approach renders analytical results provided $J(0)$ and $dJ/dm_f^2(0)$ can be computed analytically. In such case, their values can be used as the boundary conditions to find the solution. It is straightforward to show that indeed this is the case and that
\vspace{1.2cm}
\begin{eqnarray}
J(0)&=&\frac{T^4}{\pi^2}\left[\text{Li}_4\left(-e^{-\frac{\mu}{T}}\right)+\text{Li}_4\left(-e^{\frac{\mu}{T}}\right)\right],\nonumber\\
\frac{dJ}{dm_f^2}(0)&=&-\frac{T^2}{2\pi^2}\left[\text{Li}_2\left(-e^{-\frac{\mu}{T}}\right)+\text{Li}_2\left(-e^{\frac{\mu}{T}}\right)\right],
\label{firstJ}
\end{eqnarray}
where Li$_{n}$ is the polylogarithmic function of order $n$. The approximation of $d^2J/(dm_f^2)^2$ up to $\mathcal{O}(m_f^2)$ is found using the same procedure employed to find Eq.~(\ref{casifinfer}) and the result is
\begin{eqnarray}
\frac{d^2J}{(dm_f^2)^2}&=&\frac{1}{8\pi^2}\Big[\ln\left(\frac{m_f^2}{T^2}\right)+ \psi^{0}\left(\frac{3}{2}\right) \nonumber\\
&-& \psi^{0}\left(\frac{1}{2}+\frac{i\mu}{2\pi T}\right)- \psi^0\left(\frac{1}{2}-\frac{i\mu}{2\pi T}\right)\nonumber\\
&-&2\left(1+\ln(2\pi)\right)+\gamma_E\Big].
\label{secondferterm}
\end{eqnarray}
We now integrate Eq.~(\ref{secondferterm}) considering it as a differential equation and implementing Eqs.~(\ref{firstJ}) as the boundary conditions for the first derivative and the function itself. Integrating Eq.~(\ref{secondferterm}) once, we obtain
\begin{eqnarray}
J(m_f^2)&=&\frac{m_f^4}{16\pi^2}\Big[\ln\left(\frac{m_f^2}{T^2}\right)-\frac{3}{2}+ \psi^{0}\left(\frac{3}{2}\right)\nonumber\\
&-& \psi^{0}\left(\frac{1}{2}+\frac{i\mu}{2\pi T}\right)- \psi^{0}\left(\frac{1}{2}-\frac{i\mu}{2\pi T}\right)\nonumber\\
&-&2\left(1+\ln(2\pi)\right)+\gamma_E\Big] +C_1m_f^2 + C_2.
\label{int}
\end{eqnarray}
Using the set of Eqs.~(\ref{secondferterm}), we infer that
\begin{eqnarray}
C_1&=&-\frac{T^2}{2\pi^2}\left[\text{Li}_2\left(-e^{-\frac{\mu}{T}}\right)+\text{Li}_2\left(-e^{\frac{\mu}{T}}\right)\right],\nonumber\\
C_2&=&\frac{T^4}{\pi^2}\left[\text{Li}_4\left(-e^{-\frac{\mu}{T}}\right)+\text{Li}_4\left(-e^{\frac{\mu}{T}}\right)\right].
\label{C2}
\end{eqnarray}
Therefore, writing Eqs.~(\ref{ferrenvac}),~(\ref{casifinfer}), and~(\ref{int}) together, the contribution from one fermion species to the one-loop effective potential is

\begin{widetext}
\begin{eqnarray}
V_f^{(\text{1;B ren})}+V_f^{(\text{1;B matt})}&=&\frac{m_f^4}{16\pi^2}\Bigg[ \ln{\bigg(\frac{\tilde{\mu}^2}{m_f^2}\bigg)}+\frac{3}{2} \Bigg]+\frac{|q_fB|^2}{24\pi^2}\Bigg[ \ln{\bigg(\frac{\tilde{\mu}^2}{m_f^2}\bigg)} \Bigg] \nonumber \\
&+&\frac{m_f^4}{16\pi^2}\Bigg[ \ln{\bigg( \frac{m_f^2}{T^2}\bigg)-\frac{3}{2}+\psi^0\bigg( \frac{3}{2}  \bigg)}\nonumber \\
&-&\psi^0\bigg( \frac{1}{2}+\frac{i\mu}{2\pi T} \bigg)-\psi^0\bigg( \frac{1}{2}-\frac{i\mu}{2\pi T} \bigg) -2(1+\ln{(2\pi)})+\gamma_E \Bigg]\nonumber\\
&-&\frac{m_f^2 T^2}{2\pi^2}\Big[ \text{Li}_2(-e^{-\frac{\mu}{T}})+\text{Li}_2(-e^{\frac{\mu}{T}}) \Big] \nonumber \\
&-&\frac{|q_fB|^2}{12\pi^2}\Bigg\{ 2\ln{(2\pi)}-\gamma_E -\frac{1}{2}\psi^0\bigg( \frac{1}{2} \bigg) \nonumber\\
&-& \frac{1}{2}\psi^0\bigg( \frac{1}{2}+\frac{i\mu}{2\pi T} \bigg)- \frac{1}{2}\psi^0\bigg( \frac{1}{2}-\frac{i\mu}{2\pi T} \bigg)+\frac{1}{2}\ln{\bigg( \frac{4\pi^2 T^2}{m_f^2} \bigg)}  \nonumber\\
&+&\frac{m_f^2}{16\pi^2 T^2}\bigg[ \zeta\bigg( 3,\frac{1}{2}+\frac{i\mu}{2\pi T} \bigg)+ \zeta\bigg( 3,\frac{1}{2}-\frac{i\mu}{2\pi T} \bigg)\bigg]  \Bigg\}.
\label{feronelooptot}
\end{eqnarray}
Equation~(\ref{feronelooptot}) possesses the remarkable property that potentially offending logarithmic terms, as the fermion mass vanishes, combine to trade the mass dependence by a temperature dependence. Thus, bringing together these terms and simplifying, we obtain
\begin{eqnarray}
V_{f}^{(eff)}\equiv  V_{f}^{(1;B\ ren)}+V_{f}^{(1;B\ matt)}&=&\frac{m_f^4}{16\pi^2}\left[\ln\left(\frac{\tilde{\mu}^2}{T^2}\right)
\right]+\frac{|q_fB|^2}{24\pi^2}\left[\ln\left(\frac{\tilde{\mu}^2}{4\pi^2T^2}\right)\right]\nonumber\\
&+&\frac{m_f^4}{16\pi^2}\Big[ \psi^{0}\left(\frac{3}{2}\right)-2\left(1+\ln(2\pi)\right)+\gamma_E\nonumber\\
&-& \psi^{0}\left(\frac{1}{2}+\frac{i\mu}{2\pi T}\right)- \psi^{0}\left(\frac{1}{2}-\frac{i\mu}{2\pi T}\right)\Big]\nonumber\\ 
&-&\frac{m_f^2T^2}{2\pi^2}\left[\text{Li}_2\left(-e^{-\frac{\mu}{T}}\right)+\text{Li}_2\left(-e^{\frac{\mu}{T}}\right)\right]\nonumber\\ 
&+&\frac{T^4}{\pi^2}\left[\text{Li}_4\left(-e^{-\frac{\mu}{T}}\right)+\text{Li}_4\left(-e^{\frac{\mu}{T}}\right)\right]\nonumber\\
&-&\frac{|q_fB|^2}{12\pi^2}\left\{2\ln(2\pi)-\gamma_E-\frac{1}{2} \psi^0\left(\frac{1}{2}\right)\right.\nonumber\\
&-&\frac{1}{2} \psi^0\left(\frac{1}{2}+\frac{i\mu}{2\pi T}\right)-\frac{1}{2} \psi^0\left(\frac{1}{2}-\frac{i\mu}{2\pi T}\right)\nonumber\\
&+&\left.\frac{m_f^2}{16\pi^2T^2}\left[\zeta\left(3,\frac{1}{2}+\frac{i\mu}{2\pi T}\right)+\zeta\left(3,\frac{1}{2}-\frac{i\mu}{2\pi T}\right)\right]\right\}.\nonumber \\
\label{feronelooptotfin}
\end{eqnarray}
\end{widetext}
Before writing together the fermion and boson contributions to the effective potential, it is necessary to discuss the conditions that need to be implemented to avoid that the inclusion of the one-loop corrections distort the vacuum of the theory from where the particle masses are defined. These are the vacuum stability conditions.

\subsection{Vacuum stability}\label{secII-4}

In order to complete the description of the effective potential, it is important to notice that the vacuum, one-loop $B$-independent corrections, distort the original tree-level potential. This distortion manifest itself in a shift of the minimum and its curvature in the $\sigma$-direction. When ignored, this distortion leads to changes in what we identify as the vacuum $\sigma$, pion and fermion masses. However it is clear that our description of the vacuum at any order in perturbation theory should be consistent with the measured vacuum particle masses. This means that the vacuum distortion needs to be compensated in such a way that its properties, directly related to particle masses, keep being the same as if these properties were defined from the tree-level potential. This procedure is dubbed {\it vacuum stabilization}~\cite{Carrington:1991hz}. When vacuum stability is not considered, the description of the nature of the phase transition can lead to drastically different results such as the possible splitting of the deconfinement and chiral transitions in an external
magnetic field. This scenario has been studied in Ref.~\cite{Mizher:2010zb} within the LSMq
coupled to a Polyakov loop. The authors found that the vacuum correction from
quarks on the phase structure was dramatic. When ignoring this correction, the confinement and
chiral phase transition lines coincide. However, inclusion of the correction led to a splitting of
the confinement and chiral transition lines, and both chiral and deconfining critical temperatures
became increasing functions of the magnetic field. The vacuum contribution from the quarks
drastically affected the chiral sector as well. Since these conclusions were drawn from a model
that does not reproduce the behavior of the critical temperature as a function of the magnetic
field found by lattice QCD calculations, they are nowadays regarded as incorrect. Furthermore,
LQCD results shows that no significant difference between chiral and deconfinement
transition temperatures exists up to fields as intense as 3.25 GeV$^2$~\cite{Endrodi:2015oba}.
\begin{figure}[b]
\centering
\includegraphics[scale=0.53]{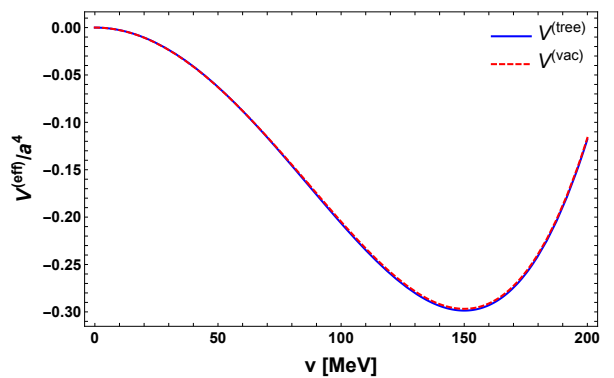}
\caption{Comparison between the tree-level and stabilized vacua. Notice that inclusion of the vacuum stability conditions produces that both terms coincide around the minimum.}
\label{vacstab}
\end{figure}
In order to stabilize the vacuum at one-loop level, we look at the tree-level plus one-loop $B$-independent fermion and boson vacua. For the latter we only include the potential in the direction of the $\sigma$-field. These contributions are given by
\begin{eqnarray}
V^{(vac)}&\equiv& V^{(tree)}+V_\sigma^{(1)}+V_f^{(1)}\nonumber\\
&=&-\frac{a^{2}}{2}v^{2} +\frac{\lambda}{4}v^{4}-\frac{\delta a^2}{2}v^2+\frac{\delta\lambda}{4}v^4\nonumber\\
&+&\frac{m_\sigma^4}{64\pi^2}\left[\ln\left(\frac{m_\sigma^2}{\tilde{\mu}^2}\right)-\frac{3}{2}
\right]\nonumber\\
&-&N_cN_f\frac{m_f^4}{16\pi^2}\left[\ln\left(\frac{m_f^2}{\tilde{\mu}^2}\right)-\frac{3}{2}
\right],
\label{vacuumstab}
\end{eqnarray}
where we have accounted for the contribution of $N_cN_f=6$ fermions, corresponding to $N_c$ colors and $N_f$ flavors and have introduced the {\it counterterms} $\delta a^2$ and $\delta\lambda$. These counterterms need to be fixed from the conditions to keep the vacuum and its curvature at their tree-value levels, namely
\begin{eqnarray}
\frac{1}{v}\frac{dV^{(vac)}}{dv}\Big|_{v=v_0}=0,\nonumber\\
\frac{d^2V^{(vac)}}{dv^2}\Big|_{v=v_0}=2a^2,
\label{stability}
\end{eqnarray}
where $v_0=\sqrt{a^2/\lambda}$ is the value of $v$ at the tree-level minimum. The solution for the counterterms $\delta a^2$ and $\delta\lambda$ are given by
\begin{eqnarray}
\delta a^2&=&-\frac{3a^2}{16\pi^2}\left[\lambda\ln\left(\frac{2a^2}{\tilde{\mu}^2}\right)-8\frac{g^4}{\lambda}+2\lambda\right],\nonumber\\
\delta\lambda&=&\frac{3}{16\pi^2}\left[8g^4\ln\left(\frac{g^2a^2}{\lambda\tilde{\mu}^2}\right)-3\lambda^2\ln\left(\frac{2a^2}{\tilde{\mu}^2}\right)\right].
\label{counter}
\end{eqnarray}
Figure~\ref{vacstab} shows the three level potential $V^{(tree)}$ compared to the vacuum potential $V^{(vac)}$ after implementing the stabilization procedure. Notice that both coincide near the minimum.

\begin{figure*}[t]
\begin{center}
\includegraphics[scale=1]{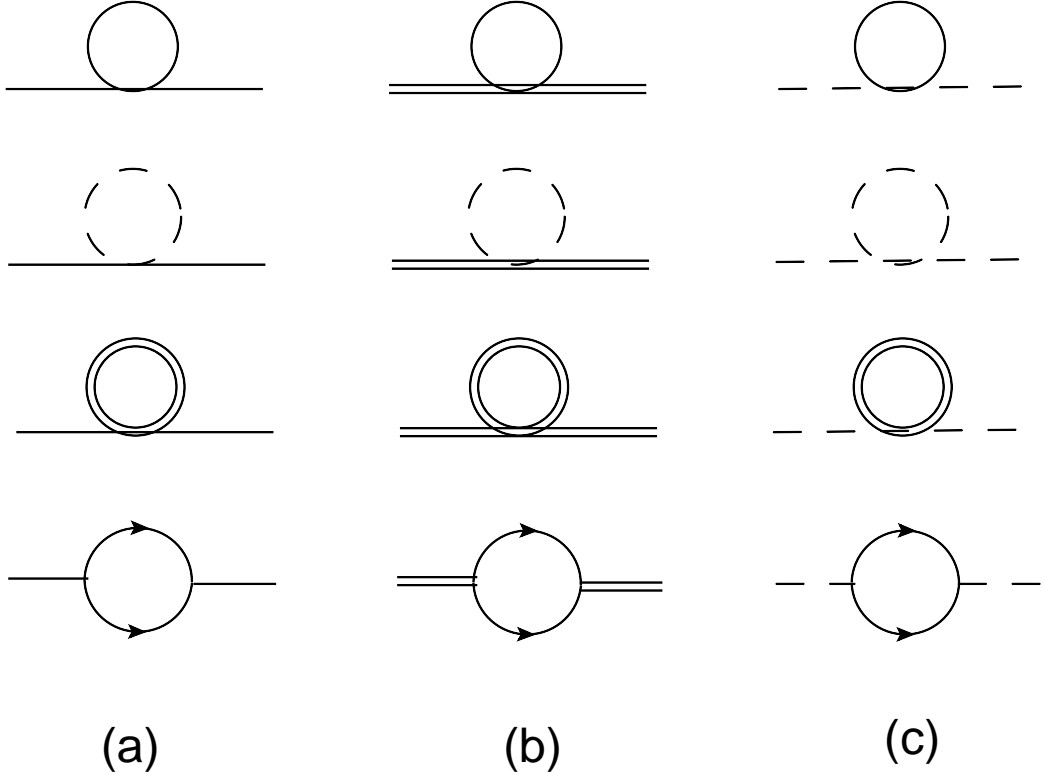}
\end{center}
\caption{Feynman diagrams contributing to the one-loop boson self-energies. Each column represents the diagrams contributing to the self-energy of a given boson: (a) the sigma, (b) the the neutral pion and (c) the charged pions. The continuous lines represent the sigma, the double lines the neutral pion, the dashed lines the charged pions and the continuous lines with arrows the quarks $u$ and $d$.}
\label{fig2}
\end{figure*}
Writing all the ingredients together, the effective potential up to the ring diagrams contribution and after vacuum stabilization can be written as

\begin{widetext}
\begin{eqnarray}
V^{(eff)}&=&-\frac{a^{2}}{2}\left\{1+\frac{3a^2}{8\pi^2}\left[\lambda\ln\left(\frac{2a^2}{\tilde{\mu}^2}\right)-8\frac{g^4}{\lambda}+2\lambda\right]\right\}v^{2}\nonumber\\ &+&\frac{\lambda}{4}\left\{1+\frac{3}{4\pi^2}\left[8g^4\ln\left(\frac{g^2a^2}{\lambda\tilde{\mu}^2}\right)-3\lambda^2\ln\left(\frac{2a^2}{\tilde{\mu}^2}\right)\right]\right\}v^{4}\nonumber\\
&+&\sum_{b=\pi^\pm,\pi^0,\sigma}\left\{-\frac{T^4\pi^2}{90}+\frac{T^2m_b^2}{24}-\frac{T(m_b^2+\Pi_b)^{3/2}}{12\pi}-\frac{m_b^4}{64\pi^2}\left[\ln\left(\frac{\tilde{\mu}^2}{(4\pi T)^2}\right)+2\gamma_E\right]\right\}\nonumber\\
&-&\frac{|q_bB|^2}{24\pi^2}\sum_{b=\pi^\pm}\left\{\frac{T\pi}{2(m_b^2+\Pi_b)^{1/2}}+\frac{1}{4}\ln\left(\frac{\tilde{\mu}^2}{(4\pi T)^2}\right)+\frac{1}{2}\gamma_E\right.\nonumber\\
&-&\left.\frac{1}{4}\zeta(3)\left(\frac{m_b^2}{(2\pi T)^2}\right)+\frac{3}{16}\zeta(5)\left(\frac{m_b^4}{(2\pi T)^4}\right)\right\}\nonumber\\
&+&N_cN_f\left\{\frac{m_f^4}{16\pi^2}\Big[\ln\left(\frac{\tilde{\mu}^2}{T^2}\right)- \psi^{0}\left(\frac{1}{2}+\frac{i\mu}{2\pi T}\right)- \psi^{0}\left(\frac{1}{2}-\frac{i\mu}{2\pi T}\right)\right.\nonumber\\
&+& \psi^{0}\left(\frac{3}{2}\right)-2\left(1+\ln(2\pi)\right)+\gamma_E\Big]\nonumber\\
&-&\frac{m_f^2T^2}{2\pi^2}\left[\text{Li}_2\left(-e^{-\frac{\mu}{T}}\right)+\text{Li}_2\left(-e^{\frac{\mu}{T}}\right)\right]+\frac{T^4}{\pi^2}\left[\text{Li}_4\left(-e^{-\frac{\mu}{T}}\right)+\text{Li}_4\left(-e^{\frac{\mu}{T}}\right)\right]\nonumber\\
&+&\frac{|q_fB|^2}{12\pi^2}\left(\frac{1}{2}\ln\left(\frac{\tilde{\mu}^2}{4\pi^2T^2}\right)+\frac{1}{2} \psi^0\left(\frac{1}{2}+\frac{i\mu}{2\pi T}\right)+\frac{1}{2} \psi^0\left(\frac{1}{2}-\frac{i\mu}{2\pi T}\right)\right.\nonumber\\
&+&\left.\left.\frac{1}{2} \psi^0\left(\frac{1}{2}\right)-2\ln(2\pi)+\gamma_E-\frac{m_f^2}{16\pi^2T^2}\left[\zeta\left(3,\frac{1}{2}+\frac{i\mu}{2\pi T}\right)+\zeta\left(3,\frac{1}{2}-\frac{i\mu}{2\pi T}\right)\right]\right)\right\}, \nonumber \\
\label{vefftot}
\end{eqnarray}
\end{widetext}
where the magnetic field dependent contribution from the charged bosons has been singled out since these are the only bosons affected by the  magnetic field. In addition, we point out that, after vacuum stabilization, the position and curvature of the minimum in Eq.~(\ref{vefftot}) becomes independent of the ultraviolet renormalization scale $\tilde{\mu}$~\cite{Ayala:2020dxs}. Therefore, as long as the latter is larger than the largest of the other energy scales, one can use any value for $\tilde{\mu}$.

\subsection{Boson self-energies}

The diagrams representing the bosons self-energies are depicted in Fig.~\ref{fig2}. Each boson self-energy is made out of two distinct kinds of terms: one corresponds to the sum of boson loops and the other one to a fermion anti-fermion loop. The boson loops contribution to each boson self energy is given by
\begin{eqnarray}
\Pi_\sigma^b &=&\frac{\lambda}{4}\left[12\ I(m_\sigma)+4\ I(m_{\pi^0}) +8\ I(m_{\pi^\pm})\right],\nonumber \\
\Pi_{\pi^\pm}^b&=&\frac{\lambda}{4}\left[4\ I(m_\sigma)+4\ I(m_{\pi^0})+16\ I(m_{\pi^\pm})\right],\nonumber \\
\Pi_{\pi^0}^b&=&\frac{\lambda}{4}\left[4\ I(m_\sigma)+12\ I(m_{\pi^0})+8\ I(m_{\pi^\pm})\right],
\label{bosonloops}
\end{eqnarray}
where the function $I(m_b)$ is given by
\begin{eqnarray}
I(m_b)=T\sum_n\int \frac{d^3k}{(2\pi)^{3}}\Delta_{b}(i\omega_n,{\mathbf {k}};|q_bB|),
\label{funcI}
\end{eqnarray}
with $\Delta_b$ given by Eq.~(\ref{eff1loopB}). The factors on the right-hand side of Eq.~(\ref{bosonloops}) correspond to the combinatorial factors obtained from the interaction Lagrangian in Eq.~(\ref{lagranint}). Notice that when the propagator refers to the neutral bosons, the corresponding magnetic field dependent piece will be absent. 

In order to compute Eq.~(\ref{funcI}), we notice the relation between the boson contribution to the effective potential and the  corresponding function $I(m_b)$, given by Eqs.~(\ref{eff1loopB}) and~(\ref{using}), which can be written as
\begin{eqnarray}
I(m_b)=2\frac{dV_b^{(1;B)}}{dm_b^2}.
\label{relselfVeff}
\end{eqnarray}
To include the ring diagrams effect as well as to account already for mass renormalization,  on the right-hand side of Eq.~(\ref{relselfVeff}) we use instead Eq.~(\ref{V1loopcomplete}) and write
\begin{eqnarray}
I(m_b)=2\frac{dV_b^{(eff)}}{dm_b^2}.
\label{relselfVeff2}
\end{eqnarray}

\begin{figure}[t]
\begin{center}
\includegraphics[scale=0.6]{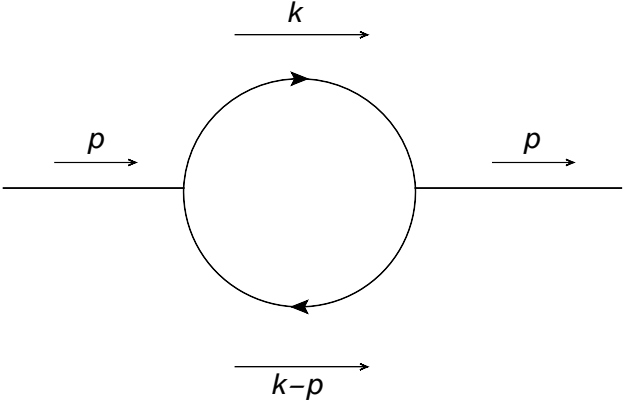}
\end{center}
\caption{Feynman diagram for the one-loop fermion contribution to the boson self-energy.}
\label{fig2-2}
\end{figure}

The fermion loop contribution to the boson self-energy, depicted in  Fig.~\ref{fig2-2} is given by
\begin{eqnarray}
\!\!\!\!\!\!\!\!\!\!\!\!\Pi^f(\omega_m,\vec{p};m_f)&=&-g^2 
\text{Tr}\left[S_f(\tilde{\omega}_n-i\mu,\vec{k};m_f)\right. \nonumber\\
&\times&\left. S_f(\tilde{\omega}_n-i\mu-\omega_m,\vec{k}-\vec{p};m_f)\right],
\label{selfEF}
\end{eqnarray}
where $S_f$ is, as before, the fermion propagator in the presence of the magnetic field and the trace refers both to the Lorentz and momentum spaces. Notice that this contribution is the same for all boson species. Equation~(\ref{selfEF}) depends on the (external) boson frequency $\omega_m$ and momentum $\vec{p}$. In order to find a suitable expression, we approximate the self-energy by its leading, momentum independent, term. This limit is found by taking  $\omega_m=\vec{p}=0$ In this limit, Eq.~(\ref{selfEF}) becomes
\begin{eqnarray}
\Pi^f(\omega_m,\vec{p};m_f)&=&-g^2 
\text{Tr}\left[S_f^2(\tilde{\omega}_n-i\mu,\vec{k};m_f) \right],
\label{selfEF2}
\end{eqnarray}
From Eq.~(\ref{further5}), we thus see that in this approximation, as in the case of the boson loop contributions, the  contribution from one fermion species to the self-energy can be obtained after mass renormalization as
\begin{eqnarray}
\Pi^f=2g^2\frac{dV_f^{(eff)}}{dm_f^2},
\label{fercontself}
\end{eqnarray}
where $V_f^{(eff)}$ is given by Eq.~(\ref{feronelooptotfin}). Therefore, the complete boson self-energies for each boson species is given by
\begin{eqnarray}
\Pi_\sigma &=&\frac{\lambda}{4}\left[12\ I(m_\sigma)+4\ I(m_{\pi^0}) +8\ I(m_{\pi^\pm})\right]+N_fN_c\Pi^f,\nonumber \\
\Pi_{\pi^\pm}&=&\frac{\lambda}{4}\left[4\ I(m_\sigma)+4\ I(m_{\pi^0})+16\ I(m_{\pi^\pm})\right]+N_fN_c\Pi^f,\nonumber \\
\Pi_{\pi^0}&=&\frac{\lambda}{4}\left[4\ I(m_\sigma)+12\ I(m_{\pi^0})+8\ I(m_{\pi^\pm})\right]+N_fN_c\Pi^f,\nonumber\\
\label{bosonselfcompl}
\end{eqnarray}

To compute the self-energy for each of the bosons, we need knowledge of the explicit form of  Eq.~(\ref{V1loopcomplete2}), given by
\begin{eqnarray}
I(m_b)&=&\frac{T^2}{12}-\frac{T}{4\pi}\left(m_b^2+\Pi_b\right)^{1/2}\nonumber\\
&-&\frac{m_b^2}{16\pi^2}\left[\ln\left(\frac{\tilde{\mu}^2}{(4\pi T)^2}\right)+2\gamma_E\right]\nonumber\\
&-&\frac{|q_bB|^2}{12\pi^2}\left[-\frac{\pi T}{4\left(m_b^2+\Pi_b\right)^{3/2}}-\frac{1}{4}\frac{\zeta(3)}{(2\pi T)^2}\right. \nonumber\\
&+&\left. \frac{3}{8}\zeta(5)\left(\frac{m_b^2}{(2\pi T)^4}\right)\right],
\label{Ibexpl}
\end{eqnarray}
where $\Pi_b$ includes the fermion contribution to the boson self-energy with flavor $b$.
Notice that when  Eq.~(\ref{Ibexpl}), is used into Eq.~(\ref{bosonselfcompl}), the resulting equations need to be solved self-consistently. However, working at high-temperature, a suitable approximation for the function $I_b$ consists of using only the leading matter contributions appearing on the right-hand side of Eq.~(\ref{Ibexpl}), writing
\begin{eqnarray}
I(m_b)&=&\frac{T^2}{12},\nonumber\\
\Pi^f&=&-g^2\frac{T^2}{\pi^2}\left[\text{Li}_2\left(-e^{-\frac{\mu}{T}}\right)+\text{Li}_2\left(-e^{\frac{\mu}{T}}\right)\right].
\label{Ibleading}
\end{eqnarray}
Since in this approximation the function $I(m_b)$ becomes independent of the boson species, we can write
\begin{eqnarray}
\Pi_1&\equiv&\Pi_\sigma=\Pi_{\pi^\pm}=\Pi_{\pi^0}\nonumber\\
&=&\lambda\frac{T^2}{2}-N_fN_cg^2\frac{T^2}{\pi^2}\left[\text{Li}_2\left(-e^{-\frac{\mu}{T}}\right)+\text{Li}_2\left(-e^{\frac{\mu}{T}}\right)\right].\nonumber\\
\label{Pileading}
\end{eqnarray}
Equation~(\ref{vefftot}), with the boson self-energies given by Eq. (\ref{Pileading}) constitutes the main tool to study the QCD phase diagram from the point of view of chiral symmetry restoration/breaking. 

In order to achieve better accuracy, it has been shown that the inclusion of thermo-magnetic  modifications of the couplings, to account for their running with temperature, density as well as with the strength of the magnetic field, is important~\cite{Ayala:2014gwa,Ayala:2014iba}. However, in the static and infrared limit, these corrections turn out to be inversely proportional to the particles masses. Thus, when working in the strict chiral limit, as we do in this work, complications to define the proper infrared regulator arise. We thus postpone the discussion of this issue for a future work and proceed to find conditions to determine the parameters of the model.

\subsection{Parameter fixing}

The LSMq contains three independent parameters, namely, the Lagrangian squared mass parameter $a^2$ and the boson and fermion-boson couplings $\lambda$ and $g$. For a complete description of the phase diagram, these parameters need to be fixed using conditions suitable for finite $T$ and $\mu$ and not from vacuum conditions. In this section, we describe the procedure to fix these parameters. Different ways to try fixing the parameters $\lambda$ and $g$ have been previously discussed but the conclusions are unclear. It is however important to mention that when conditions to fix the latter contain the possibility of a first order phase transition for $T=0$ and $\mu=m_B/3$, where $m_B\sim 1$ GeV is the typical baryon mass, the CEP turns out to be located at low values of $T$ and high values of $\mu$~\cite{Ayala:2019skg,Ayala:2017ucc}.

\begin{figure}[t]
\centering
\includegraphics[scale=0.53]{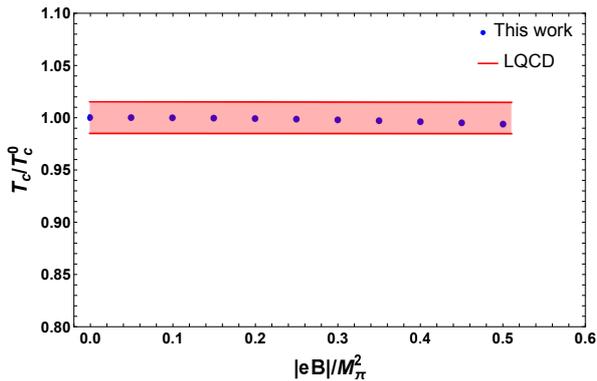}
\caption{Critical temperature as a function of the magnetic field strength in units of the pion mass, $M_\pi=140$ MeV, squared. For the set of parameters $\lambda=1.6$, $g=0.794$ and $a=133.538$ MeV, and $\mu=0$ for the case when the effective potential is computed up to the contribution of the ring diagrams. The blue dots indicate a second order phase transition.}
\label{figIMCm100}
\end{figure}

Recall that the boson self-energy represents the thermo-magnetic correction to the boson mass. For a second order (our proxy for a crossover), namely, a continuous phase transition, these corrections should produce that the thermal boson masses vanish when the symmetry is restored. This means that at the phase transition, the effective potential develops not only a minimum but it is also flat (the second derivative vanishes) at $v=0$. This property can be exploited to find the suitable value of $a$ at the critical temperature $T_c$ for $\mu=0$. Since the thermal boson masses are degenerate at $v=0$, from Eqs.~(\ref{bosonselfcompl}) and~(\ref{Ibexpl}) the condition $\Pi-a^2=0$ can be written as
\begin{eqnarray}
6&\lambda&\left(\frac{T_c^2}{12}-\frac{T_c}{4\pi}\left(\Pi_1-a^2\right)^{1/2}\right. \nonumber\\
&+&\left. \frac{a^2}{16\pi^2}\left[\ln\left(\frac{\tilde{\mu}^2}{(4\pi T_c)^2}\right)+2\gamma_E\right]\right)\nonumber\\
&+&g^2T_c^2-a^2=0.
\label{condatTc}
\end{eqnarray}
where from Eq.~(\ref{Pileading})
\begin{eqnarray}
\Pi_1(T_c,\mu=0)=\left[\frac{\lambda}{2} + g^2\right]T_c^2.
\label{Pi1formu0}
\end{eqnarray}
Since we have a relative freedom to chose the value of $\tilde{\mu}$, we take $\tilde{\mu}=500$ MeV. This value is chosen large enough so as to consider it the largest scale in the problem. Since the dependence of the result is only logarithmic in $\tilde{\mu}$, variations in this parameter do not affect significantly the result. Since the values of $\lambda$, $g$ and $a$ we are looking for need to be computed including effects at finite $T$ and $\mu$, we look for solutions of Eq.~(\ref{condatTc}) using input form the LQCD transition curve for $\mu\simeq 0$. 

Using the LQCD findings for $T_c\simeq 158$ MeV and the curvature parameters $\kappa_2$ and $\kappa_4$~\cite{PhysRevLett.125.052001}, we can compute the values of $\lambda$, $g$ and $a$ that best describe the transition curve near $\mu\sim 0$, using Eq.~(\ref{condatTc}). Since this is a non-linear equation, the solution is not unique. We hereby show results for two possible sets of values and reserve the discussion of the complete search and properties of the parameter space for a future work.

\subsection{Inverse magnetic catalysis and the magnetized phase diagram in the LSMq}\label{secII-g}

\begin{figure}[t]
\centering
\includegraphics[scale=0.53]{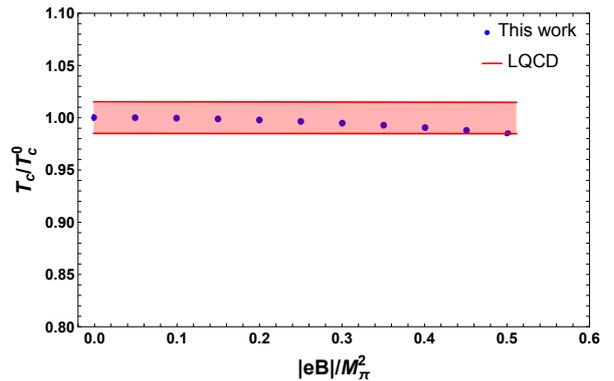}
\caption{Critical temperature as a function of the magnetic field strength in units of the pion mass, $M_\pi=140$ MeV, squared for the set of parameters $\lambda=2$, $g=0.484$ and $a=80.568$ MeV, and $\mu=0$ for the case when the effective potential is computed up to the contribution of the ring diagrams. The blue dots indicate a second order phase transition.}
\label{figIMCm120}
\end{figure}

Armed with the effective potential up to the ring diagram contributions, and with the parameters fixed from information on the LQCD critical curve near $\mu=0$, we can now explore the consequences for chiral symmetry restoration.

First we study whether the model describes IMC. Figures~\ref{figIMCm100} and~\ref{figIMCm120} show the critical temperature as a function of the field strength in units of the pion mass $M_\pi=140$ MeV squared, $|eB|/M_\pi^2$, normalized to the critical temperature at $\mu=0$ using two sets of parameters: $\lambda=1.6$, $g=0.794$, $a=133.538$ MeV and $\lambda=2$, $g=0.484$ $a=80.568$ MeV, respectively. The figures also show a comparison with the behavior of the critical temperature as a function of the field strength using LQCD data from Ref.~\cite{Bali:2011qj} (red band). Notice that in both cases, the critical temperature decreases as a function of the field strength showing that the magnetic field corrections produce IMC. The decrease is slightly weaker for the first set of parameters. Since the computation is valid in the weak field limit, we have restricted the magnetic field values to cover a range up to $|eB|/M_\pi^2=0.5$. Even though this range is limited, it already shows the trend whereby the critical temperature decreases as a function of the field strength.
\begin{figure}[t]
\centering
\includegraphics[scale=0.53]{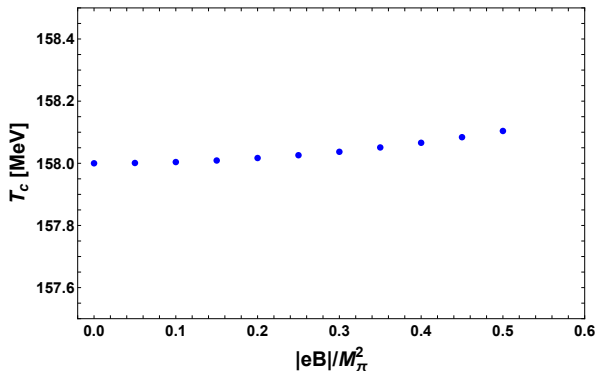}
\caption{Critical temperature as a function of the magnetic field strength in units of the pion mass $M_\pi=140$ MeV squared, for the set of parameters $\lambda=2$, $g=0.484$ and $a=80.568$ MeV and $\mu=0$ for the case when the effective potential does not contain the contribution of the ring diagrams. All the transitions are second order.}
\label{norings}
\end{figure}

In order to check whether the inclusion of the ring diagram contribution can be linked to IMC, we artificially remove these terms from the effective potential, both in the thermal as well as in the thermo-magnetic parts. Figure~\ref{norings} shows the critical temperature as a function of the magnetic field strength in units of the pion mass $M_\pi=140$ MeV squared for $\lambda=2$, $g=0.484$ and $a=80.568$ MeV and $\mu=0$. Notice that without the inclusion of the ring diagrams in the analysis, the critical temperature shows instead a modest increase with the field strength.

\begin{figure}[t]
\centering
\includegraphics[scale=0.53]{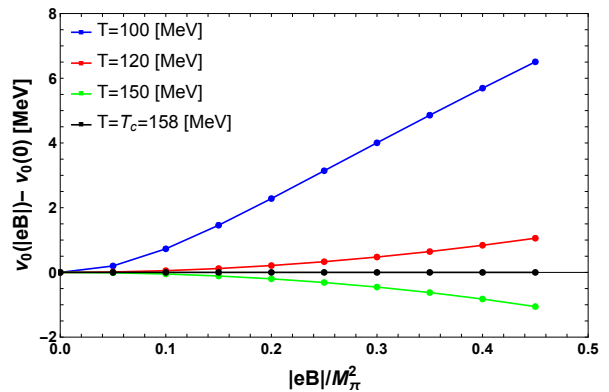}
\caption{The difference $v_0(|eB|)-v_0(0)$ as a function of the field strength in units of the pion mass $M_\pi$ squared for different temperatures using the parameters found when using $\lambda=1.6$, $g=0.794$ and $a=133.538$ MeV.}
\label{condensate}
\end{figure}

We now study the behavior of the condensate $v_0(|eB|)$ as a function of the field strength for different temperatures up to the transition temperature. Figure~\ref{condensate} shows the difference $v_0(|eB|)-v_0(0)$ as a function of the field strength in units of the pion mass $M_\pi$ squared for different temperatures using the set of parameters $\lambda=1.6$, $g=0.794$ and $a=133.538$ MeV. Notice that when the temperature is well below the critical temperature $T_c=158$ MeV, the condensate grows. However, when the temperature approaches $T_c$, this growth is tamed such that when the temperature is close to $T_c$ the growth turns into a decrease. Since we are working in the strict chiral limit, for temperatures equal or above $T_c$, the condensate reaches its equilibrium value at the symmetry restored phase $v_0(|eB|)=0$. In contrast with the analysis for the behavior of the critical temperature as a function of the field strength, in this case a comparison with LQCD data is not possible given that the first LQCD point is well above the range of field strengths that can be studied in the weak field limit. However, these findings already show the trend whereby starting from small magnetic fields the condensate increases or decreases from its vacuum value depending on whether the temperature is below or above the critical temperature in vacuum. These findings are in agreement with the general trend found by LQCD calculations~\cite{Bali:2012zg}.

\begin{figure*}[ht]
\centering
\includegraphics[scale=0.8]{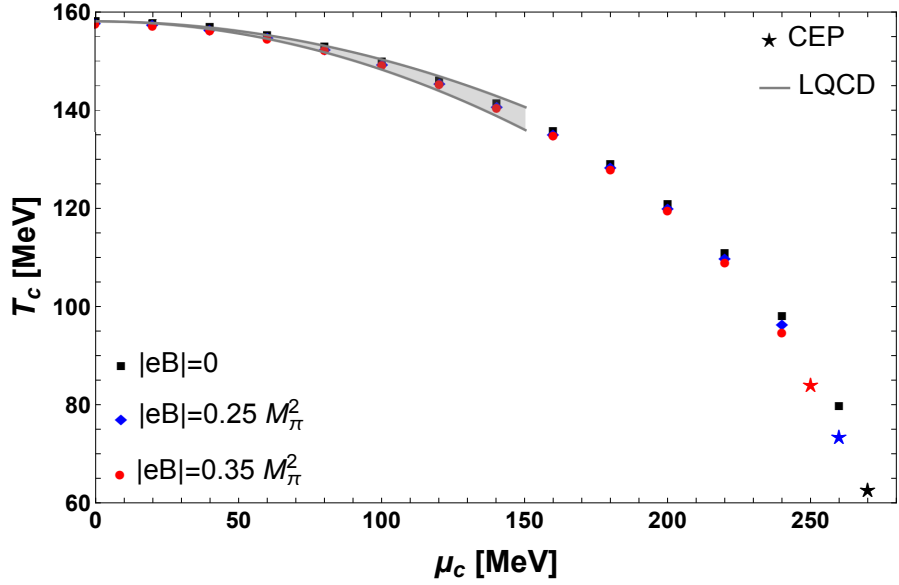}
\caption{Phase transition lines for different values of the field strength using the set of parameters $\lambda=1.6$, $a=0.794$ and $a=133.538$ MeV. The star symbols represent the position of the CEP for each value of the magnetic field.}
\label{phasediag1}
\end{figure*}

We now study the consequences for the phase diagram. Figure~\ref{phasediag1} shows the phase transition lines in the $T$--$\mu$ plane for different values of the field strength using the set of parameters $\lambda=1.6$, $g=0.794$ and $a=133.538$ MeV. Notice that as the field strength increases, the transition curves move to lower temperature values. Also, the position of the CEP found at $|eB|=0$, moves to lower values of $(\mu_c^{CEP},T_c^{CEP})$ as the field strength increases. Thus, the presence of a finite, albeit small magnetic field, produces a noticeable displacement of the CEP. For lower temperatures one needs to resort to a low--$T$ expansion to make a thorough exploration of the phase diagram~\cite{Ayala:2017ucc,Ayala:2019skg}.

We conclude this section by pointing out that the features hereby described can become more accurate, particularly when the calculation is dome for large magnetic field strengths, once the thermo-magnetic corrections to the boson self-coupling and fermion-boson coupling are included~\cite{Ayala:2020dxs} as well as when the symmetry is explicitly broken by a finite pion mass $M_\pi$.

\section{The Nambu--Jona-Lasinio model}\label{secIII}

One of the most versatile models used in the study of phase transitions in QCD is without doubt the NJL. In its more basic form, the model describes the dynamics of deconfined quarks at relatively low energy and corresponds to  a simplified version of the Schwinger-Dyson equations with integrated gluon degrees of freedom and considering only the leading interactions terms. As a result, the model consists of a non-renormalizable Dirac theory with a four-quark interaction term. 
This description of deconfined constituent quarks can be improved when including confinement effects.

The versatility of the NJL model is expressed by the fact that external sources can be incorporated in a simple way, preserving or breaking the corresponding symmetries. 
In particular, the incorporation of different chemical potentials, without the LQCD sign problem, as well as the introduction of different order parameters, resorting to a mean field approach, makes the NJL a prime tool when dealing with low-energy QCD that can also account for medium effects.

On the other hand, being non-renormalizable, the NJL model is strongly dependent on the cutoff that needs to be introduced to regulate the ultraviolet. 
In this sense, the model involves a non-unique  {\it prescription} to include the momentum cuttoff. This problem becomes more important when medium effects are considered.
Notwithstanding, this approach opens up the possibility to employ many interesting techniques as well as to use different ans\"{a}tze, according to the system that one needs to describe.
In general the cutoff is fixed in order to fit known physical quantities. 
However some care must be taken since there are critical values for the set of parameters where the solution of the gap equation turns chaotic \cite{Ahmad:2018grh,Martinez:2019ift}.

Here we analyze, from the perspective of  the NJL model, the behavior of the QCD phase diagram. In particular, we study the evolution of the CEP as a function of temperature
and baryon chemical potential, including also  the dependence on the strength of an external magnetic field.
The behavior of the CEP represents a challenge for our understanding of QCD. Qualitatively, one expects that the phase transitions describing chiral symmetry restoration and deconfinement take place at approximately the same temperature.

There are many other interesting phenomena that we will hereby not address, like cold quark matter and compact star phenomenology, where NJL models play an important role in the determination of the phase diagram in dense magnetized matter, BEC-BCS or color superconductivity \cite{Duarte:2015ppa,Coppola:2017edn}, the role of the quark anomalous magnetic moment on the thermodynamic properties of mesons immersed in a magnetic field~\cite{Chaudhuri:2020lga} or the magnetic field effects on electromagnetic probes in heavy-ion collisions such as the dilepton production rate~\cite{Islam:2018sog}. We will restrict the discussion to the description of the simultaneous temperature, baryon chemical potential and magnetic field dependence of the chiral CEP. Isospin effects for a magnetic field-induced CEP are studied in Ref.~\cite{Rechenberger:2016gim}. Multiple CEPs in magnetized three flavor quark matter, prompted by the strange sector, are studied in Ref.~\cite{Ferreira:2017wtx}. Possible signatures of the presence of a CEP in magnetized quark matter are provided studying fluctuations of conserved charges in the context of heavy-ion collisions in Ref.~\cite{Ferreira:2018pux}.
For a broader picture on other interesting aspects of the properties of magnetized QCD described from the NJL model perspective see for instance Ref.~\cite{Cao:2021rwx}.

\subsection{NJL Lagrangian}

The general expression for the $SU(3)_f$ NJL model, including scalar and pseudoscalar interactions is given by
\begin{equation}
L_\mathrm{NJL}= \bar{\psi}(i \slashed{\partial} - \hat{m})\psi + L^{4}_\mathrm{int} + L ^{6}_\mathrm{int},
\end{equation}
where $\hat{m}$ represents the diagonal mass matrix  
\begin{eqnarray}
\hat{m} = {\mathrm{diag}}(m_{u} \; m_{d}\; m_{s}).
\end{eqnarray}
$L^{4}_\mathrm{int}$ is a four-fermions interaction term that can be local or non-local. 
The original version was constructed as a local term. 
However, there are good reasons to consider non-local versions. 
Unfortunately $L^{4}_\mathrm{int}$  has a $U_{A}(1)$ symmetry  which is not observed in nature. 
The $U_{A}(1)$ symmetry can be broken by the 't Hooft term, $L^{6}_\mathrm{int}$,  which implies a six fermions interacting term. 
$L ^{4}_\mathrm{int}$ and $L ^{6}_\mathrm{int}$ are given, respectively, by
\begin{eqnarray}
L^{4}_\mathrm{int} &=& G_{S}\sum _{a=0}^{8} \left[(\bar{ \psi}\lambda ^{a} \psi)^{2} + (\bar{ \psi}i \gamma _{5}\lambda ^{a} \psi)^{2}\right]\\
L ^{6}_\mathrm{int} &=& -K\left(\mathrm{Det} _{f}[\bar{ \psi}(1+\gamma _{5})\psi] + \mathrm{Det} _{f}[\bar{ \psi}(1-\gamma _{5}) \psi]\right),\nonumber\\
\end{eqnarray}
where $\lambda _{a}$ are the Gell-Mann matrices in the flavor space, with $\lambda_0=\sqrt{2/3}\,I$.
The numbers 4 and 6 in the previous terms  indicate the number of interacting quarks in each case.  
The 't Hooft term  can be written in a more explicit way as
\begin{eqnarray}
L^{6}_\mathrm{int} &=& \frac{-K}{4}A_{abcd}(\bar{ \psi}\lambda ^{a} \psi)\nonumber\\
&\times&[(\bar{ \psi}\lambda ^{b} \psi)(\bar{ \psi}\lambda ^{c} \psi) 
- 3(\bar{ \psi}i\gamma _{5}\lambda ^{b} q )(\bar{\psi} i \gamma _{5}\lambda ^{c} \psi)],\nonumber\\
\end{eqnarray}
where the symmetric coefficients $A _{abcd}$ are given by $A_{abcd} = \frac{1}{\!3} \epsilon _{ijk}\epsilon _{mnl}(\lambda _{a})_{im}(\lambda _{b})_{jn}(\lambda _{c})_{kl}$.

In addition to the interaction terms mentioned above, one can consider other chiral invariant operators, in particular an interaction term involving vector and axial-vector currents
\begin{equation}
    L^{4}_V=G_{V}\sum_{a=0}^{8} \left[(\bar{ \psi}\gamma_\mu\lambda ^{a} \psi)^{2} + (\bar{ \psi}\gamma_\mu\gamma _{5}\lambda^{a} \psi)^{2}\right].
\end{equation}
This term contributes only if finite density effects are present. 
It is easy to see that in the mean-field approximation, the relevant operator gives the quark density $\langle \bar\psi\gamma_\mu\psi\rangle = \langle\psi^\dag\psi\rangle\delta_{\mu 0}$ \cite{Buballa:2003qv}. 
Therefore, it is not possible to obtain $G_V$ {\it a priori} from known vacuum observables.
$G_V$ must be obtained using other in-medium phenomenological arguments.
\subsection{Confinement}

The NJL model does a good description of the dynamics involving constituent quarks. However it lacks information about confinement. This information must be added and there are different methods to include it. 
The most often used one is the inclusion of the Polyakov loop. 
Another method is to consider non-local operators, and a third one commonly used method is to consider an infrared cutoff considering the representation of the propagator in terms of a Schwinger proper-time~\cite{Ahmad:2016iez}.

The original version of the NJL model \cite{Nambu:1961tp}, including quarks, has been used for the discussion of the phase transition in the mean field approximation, where the quark condensate plays the role of the mean field for chiral symmetry restoration. For a review on this subject, see Ref.~\cite{Andersen:2014xxa}. In this version of the model, the deconfinement transition is absent. Perhaps, we should mention here that in principle these transitions are  different in nature. Chiral symmetry restoration is associated to a shift from a Nambu-Goldstone realization into a Wigner-Weyl realization. In the massless quark limit this is achieved by the vanishing of the quark condensate, the order parameter for this transition, which can also be expressed in terms of the vanishing of the pion decay constant. Deconfinement in this model can be taken into account through the Polyakov term $L(x)$, defined in terms of the trace of a Wilson loop in the Euclidean time direction
\begin{equation} 
L =\frac{1}{N_{c}}  {\mbox {Tr}} P \exp{ \left[ig\int _{0} ^\beta d\tau A_{0}(\boldsymbol{x},\tau)\right]},
\end{equation}
where $g$ is the coupling in the QCD Lagrangian, $N_{c}$ is the number of colors in the theory and $A_{0} =A^{0}_{a}T^{a}$, with $T^{a}$ the generators of $SU(N)$. $\it P$ denotes the path ordering of the integral. The definition of this object is due to
Polyakov~\cite{Polyakov:1978vu}. In the context  of gauge theories at finite temperature, this object provides an order parameter for deconfinement. 't Hooft~\cite{tHooft:1976rip} has highlighted the importance of the global symmetry $Z(N)$ in gauge theories, where $Z(N)$ is the center of the group $SU(N)$. In fact, for the QCD Lagrangian we have the usual $SU(N_{c})$ transformations $\Omega$ such that  $D_{\mu} = \Omega ^{\dagger}D_{\mu}\Omega$ and $ \psi \rightarrow \Omega ^{\dagger} \psi$ where $\psi$ are the quark fields. Of course we have $\Omega \Omega ^{\dagger} =1$ and ${\mbox{Det}}\,  \Omega =1$. If we consider a gauge transformation with a constant phase multiplying the identity 
$\Omega_{c} = e^{-i\phi}I_{N_{c}\times N_{c}}$, in order for $\Omega$ to be an element of $SU(N)_{c}$, the condition ${\mbox{Det}}\, \Omega_{c} =1$ implies $\phi = 2\pi j/N_{c}$ with $j =0 \cdots  (N-1)$.

Given the above condition, the expectation value of the Polyakov loop transforms in a non-trivial way under the center group $Z _{N_{c}}$. If $\Phi = \langle L \rangle$  we expect $\langle L \rangle=\phi_0\exp{(i\,2\pi  j/N_{c})}$, where $\phi _{0}$ is a real function which should vanish for high temperatures. In fact, the expectation value of the Polyakov loop corresponds to the ratio of the partition functions of a system with an external quark inserted in a pure gluon system, $Z_{Q}$, and the partition function of a pure gluon system $Z$. {\it i.e.}, $ Z_{Q}/Z$, from where we obtain $\Phi =\langle L \rangle = \exp{(-\beta F)}$, where $F$ is the free energy of the system with a single quark. For an external quark inserted in a gluon system, this energy becomes infinity and therefore $\Phi = 0$, which is tantamount of confinement. This means that one cannot have free color degrees of freedom in the system. On the contrary, for high temperatures, since the theory becomes essentially free, $\langle L \rangle \rightarrow 1$. There must be an intermediate temperature where $\Phi$ goes between the confined and the deconfined phases. We see that the expectation value of the Polyakov loop plays the role of an order parameter for deconfinement. Notice that, in the case of chiral symmetry restoration, the order parameter is instead the quark condensate.

Let us  now discuss the NJL model with a Polyakov term, dubbed as the PNJL model. 
Let us start with the $SU(3)_f$ general NJL Lagrangian. The idea is to take into account the vacuum expectation value of the trace of the Polyakov loop, $\Phi$.  We refer to this object as the Polyakov field. 

The idea is to build an effective potential  $U(\Phi,\bar{\Phi},T)$ such that the deconfinement transition  is included. Since $\Phi = 0$ in the confined phase, this potential should have a minimum in $\Phi =0$ for $T < T_{0}$. When $T = T _{0}$  we expect that $\Phi =0$ should correspond to a local maximum and a new minimum for $\Phi  >0$ must appear triggering the deconfinement phase transition through a breaking of the $Z _{3}$  symmetry. This minimum will move to $\Phi =1$ when the temperature grows beyond $T _{0}$. 

In the literature, several potentials satisfying these conditions have been proposed. The potential 
\begin{eqnarray}
\frac{1}{T^{4}}U(\Phi,\bar{\Phi},T) &=& -\frac{a(T)}{2}\bar{\Phi}\Phi +b(T) \ln[1 -6\bar{\Phi}\Phi \nonumber\\
&+& 4 (\bar{\Phi}^{3} + \Phi ^{3}) - 3(\bar{\Phi}\Phi)^2],
\end{eqnarray}
where $a(T) = a_{0} + a_{1}(T_0/T) + a_{2}(T_0/T)^2$ and $b(T) =  b_{3}(T_0/T)^3$ reproduces results from LQCD \cite{Roessner:2006xn}.
The parameters in this expression were fixed by comparing with LQCD data. 
They are given by: $a_{0} = 3.51$, $a_{1} = -2.47$, $a_{2} = 15.2$ and $b_{3} = -1.75$, where 
$T_{0}$ is the critical temperature for a pure gauge field theory, which has been determined by LQCD as $T_{0} = 270$\;MeV.

So far we have a NJL model for the quark sector and a  potential for the Polyakov loop. 
This potential mimics in some way the interacting gluons in  QCD . 
The coupling between them is established through the covariant derivative
$D_{\mu} = \partial _{\mu} - i A_{\mu}$, where $A _{\mu} =g A ^{a}_{\mu}\frac{\lambda ^{a}}{2}$ where $A^{a}_{\mu}$ are the color gauge fields, {\it i.e.}. the gluons. 
Normally the so called Polyakov gauge is used projecting the gauge field in the 4th (Euclidean) component and parametrizing it diagonally as  $A _{4} = \phi _{3}\lambda ^{3} + \phi _{8}\lambda ^{8}$,
giving as a result $\Phi =[ \cos(\phi _{3}/T) +1]/3$.

Certainly this model is not renormalizable and, at least in its the local version, an ultraviolet cutoff is introduced as a sharp cutoff in three momentum space $\Lambda$. Several sets of values, with small variations  for the parameters involved in the model, have been used. For example in Ref.~\cite{Hatsuda:1994pi} we find $\Lambda =  631.4$\; MeV, $m_{u} = m_{d}= 5.5$ MeV, $m_{s} = 135.7$\; MeV, $G\Lambda ^{2} = 1.835 $ and $K \Lambda^{5} =9.29$. 
See also Ref.~\cite{Avancini:2012ee}.

Another perspective is obtained from the non-local versions of the NJL model (nlNJL). 
For a recent general reference on the present status of non-local effective models see Ref.~\cite{Dumm:2021vop}. There are some advantages with respect to the traditional local version which support these type of models. For example, the inclusion of form factors, that play the role of non-local regulators, avoid dealing with the problem of a sharp ultraviolet regulator  $\Lambda$ in momentum space. The precise  value of $\Lambda$ is a sort of instability in the predictions. As a second aspect, the presence of a non-local form factor in the scalar current $j _{s}(x)$ leads to a momentum dependent quark effective mass, in agreement with results form LQCD \cite{Parappilly:2005ei}.

In the case of two flavors, the most simple  nlNJL model, without ´t Hooft terms, corresponds to the Euclidean action
\begin{equation}
S_{E} = \int d^{4}x\left\{\bar{ \psi}[ -i\slashed{\partial} + m]\psi -
G_{s}\left[(J_{S})^2 +(J^a_{P})^2\right] \right\} ,
\end{equation}
where the non-locality  is encoded in the currents
\begin{eqnarray}
 J_{S}(x) &=& \int d^4 z r(z) \bar{ \psi}\left(x +\frac{z}{2}\right) \psi\left(x - \frac{z}{2}\right),\\
J^a_{P}(x) &=& \int d^4 z r(z) \bar{ \psi}\left(x +\frac{z}{2}\right)i\gamma _{5}\lambda^a \psi\left(x - \frac{z}{2}\right),\nonumber\\
\end{eqnarray}
and where the regulator $r(z)$ carries  the effect of low energy QCD interactions. Notice that we have chosen the same form factor in both currents because of chiral symmetry. If $r(z) = \delta ^{4}(z)$ we recover the local version of the NJL model. Several form factors have been used in the literature, among which the most common one is the Gaussian regulator described in Euclidean momentum space as $r(p_E^2)=\exp(-p_E^2/\Lambda^2)$, with $\Lambda$ the cutoff already included as the regulator.

\subsection{Bosonization}

Let us continue with the $SU(2)_f$ version of NJL without confinement effects.
The most simple expression for the NJL Lagrangian that preserves chiral symmetry is
\begin{equation}
    {\cal L}_\mathrm{NJL}=\bar {\psi} i\slashed{\partial} \psi
    +G\left[ (\bar {\psi}\psi)^2+(\bar {\psi} i\gamma_5\boldsymbol{\tau} \psi)^2\right],
    \label{NJL_Lagrangian}
\end{equation}
where the quark fields contain flavor and spin indexes, $G$ is the NJL coupling, and $\boldsymbol{\tau}$ are  the $SU(2)_f$ generators.
This Lagrangian is invariant under $SU(2)_L\times SU(2)_R$ transformations.

The standard method used to obtain phase transitions for in-medium NJL is the minimization of the effective potential with respect of the different order parameters that define the existence or not of a different phase. 

There are many techniques that are used in order to extract hadronic properties from the NJL Lagrangian, however we focus only on the calculation of the effective potential in the mean-field approximation. For these purposes, we then continue with the bosonization procedure.

The four-fermion operators can be transformed into a quadratic fermion Lagrangian by the integration of a Gaussian exponential, due to the introduction of auxiliary fields. 
Then Eq.\,(\ref{NJL_Lagrangian}) for two flavors can be written as
\begin{equation}
e^{iS_\mathrm{NJL}}
= {\cal N}\int D\sigma D^3\pi ~ e^{i\tilde S_\mathrm{NJL}[\psi,\bar{\psi} ,\sigma,\pi^a]},
\end{equation}
where $S_\mathrm{NJL}=\int d^4x {\cal L}_\mathrm{NJL}$, and  with the new quadratic-fermion action being 
\begin{eqnarray}
\tilde S_\mathrm{NJL} &=&\int d^4x \left(
\bar\psi\left[i\slashed{\partial} -\sigma - i\gamma_5\pi^a\tau^a\right]\psi \right. \nonumber\\ &-&\left.\frac{1}{4G}\left[\sigma^2+(\pi^a)^2\right]\right), 
\label{tildeSNJL}
\end{eqnarray}
and where where ${\cal N }$ is some normalization constant.
With the last step in mind, the NJL generating functional can be written as the path integral of fermion fields and auxiliary fields. 
The relation between bosons and fermions comes implicitly from Eq.~(\ref{tildeSNJL}). By varying the boson fields, the equations of motion give $ \sigma = -2G\langle \bar qq\rangle$ and $\pi^a = -2G\langle \bar q i\gamma_5\tau^a q\rangle$.

Since the new Lagrangian is quadratic with respect to the fermion fields, they can be easily integrated out, and the generating functional becomes expressed only in terms of the auxiliary fields
\begin{eqnarray}
 {\cal Z} 
&=&\int D \psi D\bar {\psi} \,
 e^{i S_\mathrm{NJL}[\psi,\bar\psi]}\nonumber\\
 &=&
 {\cal N}\int D \psi D\bar{\psi} D\sigma D^3\pi\,
 e^{i\tilde S_\mathrm{NJL}[ \psi,\bar\psi,\sigma,\pi^a]}\nonumber\\
&=&
 {\cal N'}\int  D\sigma D^3\pi\,
 e^{i S_\mathrm{boson}[\sigma,\pi^a]}
 \nonumber\\&&
 \label{Zboson}
\end{eqnarray}
where the bosonic action is
\begin{eqnarray}
S_\mathrm{boson} &=&
 -i\mathrm{Tr}\ln [i\slashed{\partial} -\sigma - i\gamma_5\pi^a\tau^a]\nonumber\\
 &-&\frac{1}{4G}\int d^4x \left[\sigma^2+(\pi^a)^2\right],
 \label{Sboson}
\end{eqnarray}
where the trace stands for spin, color, flavor, and configuration space.
The generating functional is now represented only by the scalar and pseudoscalar auxiliary fields which can be recognized as the sigma and pions, respectively.  Considering an expansion around the mean field, by setting   $\sigma \to\langle\sigma\rangle +\sigma $, $\pi^a \to \langle\pi^a\rangle +\pi^a$, $\langle\sigma\rangle=\bar\sigma\neq 0$ and, in the absence pion condensation, $ \langle\pi^a\rangle =0$, chiral symmetry is spontaneously broken. 

Additionally we can break chiral symmetry  explicitly by adding a mass term to the original Lagrangian, namely, $-m\bar\psi \psi$. 
Considering the mean field approximation, we can obtain from Eq.~(\ref{Zboson}) the effective potential $S_\mathrm{boson}[\sigma,\pi]\to -V_4\Gamma(\bar\sigma)$, where the functional trace in momentum space can be written as
\begin{equation}
 \Gamma = -iN_f N_c\int\frac{d^4p}{(2\pi)^4}\mathrm{Tr}\ln(\slashed{p}-m-\bar\sigma) +\frac{\bar\sigma^2}{4G},
 \label{effPot}
\end{equation}
and where the current quark mass term was introduced and the trace is over Dirac matrices.
The value of the mean field $\bar \sigma$ is obtained from the minimum of the effective potential. 
The relation $\partial \Gamma/\partial\bar\sigma = 0$, produces the gap equation
\begin{equation}
 \frac{2\bar\sigma}{G} =N_fN_c \int\frac{d^4p}{(2\pi)^4}\frac{i(m+\bar\sigma)}{p^2-(m+\bar\sigma)^2+i\epsilon}, 
  \label{gapEq}
\end{equation}
where we introduced the time-ordered regulator.
The solution of the gap equation provides the value of $\bar\sigma$ as well as the effective constituent mass of the quarks $M=m+\bar\sigma$.

The introduction of the current quark mass allows us to obtain in a simple way the quark condensate. 
From the definition $-iV_4\langle\bar\psi\psi\rangle = \partial \ln{\cal Z}/\partial m$
\begin{equation}
 \langle\bar\psi \psi\rangle = -N_fN_c \int\frac{d^4p}{(2\pi)^4}\frac{i(m+\bar\sigma)}{p^2-(m+\bar\sigma)^2+i\epsilon}
 \label{qqEq}
\end{equation}
so we can see, by comparing Eq.~(\ref{gapEq}) with Eq.~(\ref{qqEq}), that the quark condensate is  related to the mean field $\bar\sigma $ by the relation $\langle\bar\psi\psi\rangle  =-\bar\sigma/2G$.

The effective potential in Eq.~(\ref{effPot}), as well as the gap equation in Eq.~(\ref{gapEq}), are UV-divergent. 
Since the theory is non-renormalizable, the regularization scheme provides another parameter of the model and must be fixed in terms of physical quantities. The most often used ones are the regularization scheme and the momentum cutoff. 
Here we have three parameters: the coupling constant $G$, the mean field $\bar \sigma$ and the cutoff $\Lambda$.
The gap equation fixes the value of the mean field $\bar \sigma$ in terms of $G$ and $\Lambda$, which are in turn obtained calculating the pion mass and pion decay constant and fixing their values from their experimental ones.
For a detailed description on the calculation of masses and decay constants see \cite{Buballa:2003qv}.

The different regularization schemes do not significantly change the result of the chiral condensate, however, when medium effects are considered, the situation changes.
A good procedure is to regularize the vacuum part only, whenever possible. 
In principle there is no reason to assume that the cutoff or couplings must be independent of medium effects.

\subsection{Temperature, chemical potential and magnetic field}

The finite temperature effects can be easily introduced in the NJL model through the Matsubara formalism. 
The simplicity of this formalism translates itself into some minimal replacements in the effective action, where now, $V_4\Gamma \to -i\beta V\Omega$, with $\Omega$ the thermodynamical potential and $\beta =1/T$ so that the generating functional in the mean-field approximation can be expressed as ${\cal Z} = e^{-\beta V \Omega}$.
The 0-component of the momentum in Eq.~(\ref{effPot}) must be replaced by fermionic Matsubara frequencies $p_0\to i\omega_n = i(2n+1)\pi T$ and consequently, the $p_0$ integral is replaced by the sum $\int dp_0 \to 2\pi iT\sum_n$.

To incorporate a chemical potential, we have to add to the NJL Lagrangian the Lagrange multiplier times the number density operator ${\cal L}_\mathrm{NJL} \to {\cal L}_\mathrm{NJL} +\mu \psi^\dag \psi$, which is then expressed as a shift in the momentum $\slashed{p}\to\slashed{p}+\mu\gamma_0$ and consequently, within the Matsubara formalism, produces a shift in the Matsubara frequencies as $\omega_n\to \omega_n-i\mu$. 

With all the above considerations, the thermodynamical potential reads as
\begin{eqnarray}
 \Omega &=& 4N_f N_c T\sum_{n=-\infty}^\infty \int\frac{d^3p}{(2\pi)^3}\ln \left[(\omega_n-i\mu)^2+\boldsymbol{p}^2+M^2\right]\nonumber\\
 &+&\frac{\bar\sigma^2}{4G}
 \label{Omega}
\end{eqnarray}
with a constituent mass given by $M=m+\bar\sigma$.
The gap equation can be obtained by minimizing the thermodynamical potential with respect to the mean field, $\partial\Omega/\partial\bar\sigma =0$.
Once the mean field value $\bar\sigma=\sigma^*(T,\mu)$ is  obtained, we can also get the quark number density $n_q = -\partial\Omega/\partial\mu|_{\bar\sigma=\sigma^*}$ and the pressure $P=-\Omega_\mathrm{reg}|_{\bar\sigma=\sigma^*}$, where the regularized thermodynamical potential is obtained by subtracting the medium contribution $\Omega_\mathrm{reg} = \Omega-\Omega|_{T,\mu=0}$.

The presence of an external magnetic field modifies  the NJL Lagrangian by adding the electromagnetic vector potential through the minimal coupling scheme, by setting the term $i\partial_\mu \to i\partial_\mu +e A_\mu$, where, for the case of a uniform constant field $\boldsymbol{B}=B\boldsymbol{\hat z}$, the vector potential reads as $A = (0,0,-B x,0)$ in the Landau gauge, or $A=\frac{1}{2}(0,B y,-Bx,0)$ in the symmetric gauge. 

The main methods to deal with NJL under an external magnetic field are: 1) the use of the wave function and energy spectrum of the effective Lagrangian, 2) the Schwinger proper time method and 3) the Ritus method. 
The Schwinger method provides the fermion propagator in the presence of the external field in Eq.~(\ref{Omega}).
It is possible to express the propagator in the form of a proper time integral that can be expanded in a power series of the magnetic field and  can also be expanded as a Laguerre series in terms of the Landau levels.

The three methods mentioned above provide the result in terms of a sum over Landau levels.
Since we are in a first instance calculating the thermodynamical potential at the one-loop level, this will not be affected by the Schwinger phase. As a result, the problem simplifies to the following rules: 
Once the trace is performed, the result depends only on the energy spectrum. 
The replacement of the energy function $E(\boldsymbol{p})\to E_l(p_z)$ where the transverse momentum changes as $\boldsymbol{p}_\perp^2\to 2|e_qB|l$, and the integral in momentum space also changes as $\int d^2p_\perp\to |e_fB|\sum_{l=0}^{\infty}$. The charge $e_f$ stands for the respective quark charge $e_u=2e/3$ and $e_d=-e/3$.
As a result, the thermodynamical potential can be written as 
\begin{eqnarray}
 \Omega &=& \frac{\bar\sigma^2}{4G} + 4N_c\, T\frac{|eB|}{2\pi}\sum_{n=-\infty}^{\infty}\sum_{l=0}^\infty  \sum_{f=u,d}(2-\delta_{l0}) \nonumber\\
 &\times&
\int\frac{dp_z}{2\pi}
\ln \left[(\omega_n-i\mu)^2+p_z^2+2|e_fB|l+M^2\right].
\nonumber\\
 \label{OmegaB}
\end{eqnarray}

In the case where a chemical potential is introduced, the contribution of the vector operator  $-G_V(\bar \psi\gamma_\mu \psi)^2$  in the NJL Lagrangian turns out to be relevant, because the mean field value of the vector current is related to the quark number density $\langle \psi^\dag \psi\rangle$.

\subsection{Critical end point}

The calculation of the position of the CEP using the NJL model has been worked by several authors, however this is not a simple task. 
The identification of the CEP is done basically by looking at the evolution of the susceptibilities, or else, by means of looking at the shape of the thermodynamical potential in terms of the order parameter (see Ref.~\cite{Martinez:2019bwq} for a different approach).
\begin{figure}
    \centering
    \includegraphics[scale=0.53]{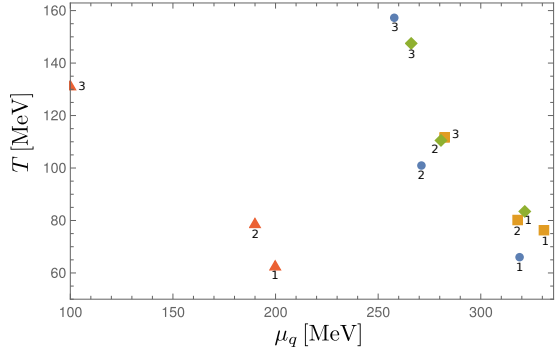}
    \caption{CEP from different authors using NJL models.\\
   Ref.\,\cite{Avancini:2012ee} (circles) with  $eB=0.195$\,GeV$^2$ (1),  $eB=0.852$\,GeV$^2$ (2) and $eB=1.95$\,GeV$^2$ (3). \\
   Ref.\,\cite{Ferrari:2012yw} (squares) with $eB=0$ (1),  $eB=0.118$\,GeV$^2$ (2), and $eB=0.294$\,GeV$^2$ (3).\\
   Ref.\,\cite{Rechenberger:2016gim} (diamonds)  $eB=0$ (1),  $eB=0.392$\,GeV$^2$ for $\langle\bar{u}u\rangle$ (2) and $\langle\bar{d}d\rangle$ (3).\\
   Ref.\,\cite{Marquez:2017uys} (triangles) with $eB=0$ (1), $eB=0.00588$\,GeV$^2$ (2) and $eB=0.0157$\,GeV$^2$ (3).}
    \label{fig:CEPNJL}
\end{figure}

Figure\,\ref{fig:CEPNJL} shows the CEP in the temperature vs. quark chemical potential ($\mu_q=\mu_B/3$) for several values of the magnetic field, using different approaches within the NJL formalism, without including the vector-current interaction. 
Basically all of them present an increment of the critical temperature as well as a reduction of the critical chemical potential as compared to calculations without a magnetic field.
In Ref.~\cite{Avancini:2012ee} (circles) the authors use a $SU(3)_f$ NJL model. 
In Ref.~\cite{Ferrari:2012yw} (squares) the authors use the $SU(2)_f$ version of NJL model.
In Ref.~\cite{Rechenberger:2016gim} (diamonds) the author considers also the $SU(2)_f$ NJL model but separating the magnetic evolution between $u$-quarks and $d$-quarks.
Finally, in Ref.~\cite{Marquez:2017uys} (triangles) the authors consider the $SU(2)_f$ non-local NJL model with a Gaussian regulator. It is interesting to observe that non-local effects change dramatically the location of the CEP position.

All the aforementioned works do not include the vector-current interaction term. 
However, this is an important contributions since its mean field is directly related with the baryon density. Also, none of the used models, in the shape they are hereby presented, describe IMC.
A better understanding of the evolution of the CEP needs to include confinement effects, IMC and  the vector-current interaction.
\begin{figure}[t]
    \centering
    \includegraphics[scale=0.53]{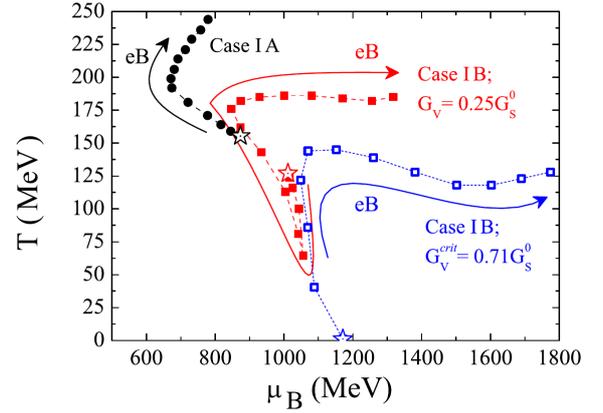}
    \caption{Magnetic evolution of the CEP from Ref.~\cite{Costa:2015bza} considering the PNJL model without IMC and a vector-current coupling $G_V$.}
    \label{fig:CEP-PCosta}
\end{figure}
Concerning the evolution of the CEP in the PNJL extended model, it is extremely interesting to stress that IMC, {\it i.e.} the fact that the critical temperature decreases as a function of the magnetic field, both for the chiral restoration and the deconfinement phase transitions, can be achieved only if the coupling $G_{s}$ acquires a dependence on the magnetic field \cite{Farias:2014eca,Ferreira:2014kpa,Ayala:2016bbi,Farias:2016gmy}. 
This dependence was first introduced in Ref.~\cite{Farias:2014eca}. Since there are no data from LQCD for the running coupling $\alpha_{s}(eB)$ that could inspire a possible $G _{s}(eB)$, a fit given by
$G_{s}(\zeta) = G_{s}^{0}(1 + a \zeta ^{2} + b\zeta ^{3})/(1 + c \zeta ^{2} + d \zeta ^{4})$, where $\zeta = eB/\Lambda _{QCD}^2$, 
and $a = 0.0108805$, $b= -1.0133  \times 10 ^{-4}$, $c = 0.02228$, $d = 1.854 \times 10 ^{-4}$, was introduced after fitting the model critical temperature for chiral restoration $T^{\chi}_{c}$ measured in LQCD at $\mu _{B}=0$~\cite{Ferreira:2014kpa}. Using this effective, magnetic field dependent  coupling in the PNJL Lagrangian, it has been shown that the critical temperatures for both transitions decrease with an increasing magnetic field strength.
\begin{figure*}[t]
    \centering
    \includegraphics{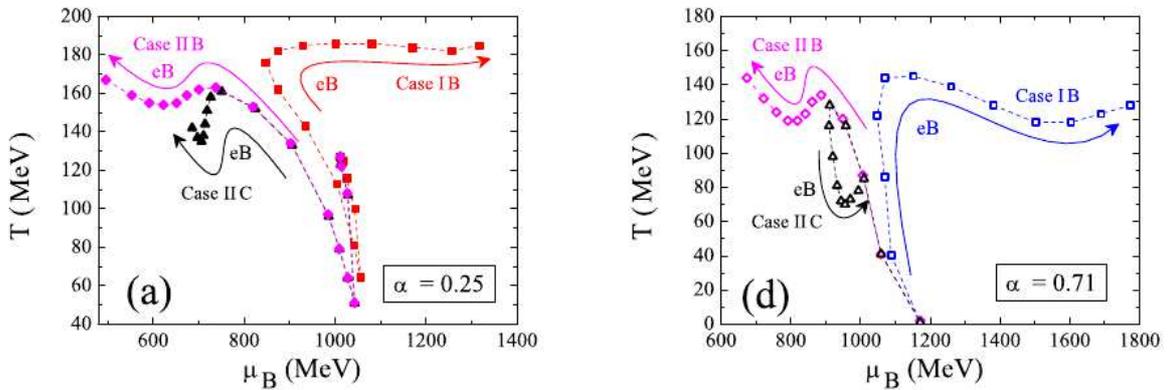}
    \caption{Magnetic evolution of the CEP from Ref.~\cite{Costa:2015bza} considering the PNJL model with IMC (cases IIB and IIC) and considering a vector-current coupling $G_V$.}
    \label{fig:CEP-PCosta_2}
\end{figure*}

For a comparison of different parametrizations of the coupling $G(T,B)$ and the conditions that must be considered see Ref.~\cite{Martinez:2018snm}.

The inclusion of the vector current coupling $-G(\bar q\gamma_\mu q)^2$ has dramatic effects on the location and evolution of the CEP. This point was discussed in Refs.~\cite{Fukushima:2008wg} and~\cite{Friesen:2014mha}. 
In fact, the CEP may even disappear, or be absent,  if the coupling $G _{v}$ becomes larger than a certain critical value $G _{v}^{crit} \approx 0.71 G _{s}^{0}$. 
Figure~\ref{fig:CEP-PCosta} shows the evolution of the CEP from $eB=0$ to $eB=1$\,GeV$^2$ considering the $SU(3)_f$ PNJL model including the vector-current coupling from Ref.~\cite{Costa:2015bza}. 
The variation of $G_V$ considerably changes the location of the CEP. 
Also there is a nontrivial evolution of the CEP, in particular when the system reaches higher values of the magnetic field, where there is a drastic inflection. 
Also the case with $G_V=0.25 G_S $ presents a non-expected behavior for low values of the magnetic field.
In all these cases IMC is absent, {\it i.e.}, when $G_{s}$ is constant.

For the case where one has an effective coupling $G_{s}(eB)$, the evolution of the CEP is different. 
Figure \ref{fig:CEP-PCosta_2} shows the evolution of the CEP with respect to the magnetic field considering IMC with $G_V=\alpha G_s(B)$ (Case IIB) as well as $G_V=\alpha G_S(0)$ (Case IIC), also compared with the case without IMC with $G_V=\alpha G_S(0)$ (Case IB).
For both values of $\alpha$ considered, the magnetic evolution of the CEP with IMC changes completely this behavior.
A more extensive analysis of the CEP considering the PNJL model with IMC effects can be founded in Ref.~\cite{Moreira:2021ety}. 
There are many open questions in the magnetic evolution of the CEP, but maybe the most important is how to obtain $G_V$. This is a non simple task and other approaches should be explored to provide insights. 

\section{Summary and perspectives}\label{concl}

In this work we have reviewed the main features of the description of the magnetized QCD phase diagram from the point of view of effective models whose main ingredient is chiral symmetry. We have focused our attention on two of these models: The LSMq and the NJL model. For the former, we have shown that a main ingredient in the description is the inclusion of plasma screening effects encoded in the resummation of the ring diagrams for the effective potential at finite temperature, baryon density and in the presence of a magnetic field. The treatment of plasma screening, that is the accounting of collective, long-wave modes, captures one of the main features near transition lines, namely, long distance correlations. This feature makes the LSMq a more powerful tool, as compared to other approaches that employ the linear sigma model in the mean field approximation, such as the one used for instance in Ref.~\cite{Andersen:2021lnk}. Inclusion of plasma screening allows to describe IMC even without the need to consider magnetic field-dependent coupling constants. Considering these effective constants helps to have a better accuracy for the description of this phenomenon. This is particularly relevant for large field strengths~\cite{Ayala:2020dxs}. The plasma screening effects are also responsible for the emergence of a CEP in the phase diagram at finite $T$ and $\mu$. The CEP moves toward lower values of $(\mu_c^{CEP},T_c^{CEP})$ even for small magnetic field strengths.

We have shown that although versatile, the NJL model is also a more limited approach in the sense that a clear separation between pure vacuum and medium effects is not always possible. Moreover, this model cannot describe IMC unless external information, such as the magnetic field dependence of the coupling, is included. The CEP identification within the NJL model is not a simple task. Calculations including non-local effects can dramatically change the CEP position. The importance of inclusion of the vector interaction has also been highlighted. The evolution of the CEP in the phase diagram also depends on whether or not the coupling includes magnetic field effects. The model can, on the other hand, incorporate in a straightforward way confinement effects by means of the coupling to the Polyakov loop or by other more rudimentary prescriptions such as the inclusion of an infrared regulator.
 
Overall, it is our impression that the LSMq and NJL model approaches provide sensible tools to explore the properties of magnetized, strongly interacting matter. However, a much needed cross talk among these methods is called for as well as a consistent physical approach to determine the model parameters from restrictive conditions so as to avoid the large dispersion in the predictions of the CEP position. In addition, calculations using the LSMq are thoroughly worked out in the weak field limit whereas in the case of the NJL models equivalent calculations are performed for strong fields. A possible avenue of encounter can be to work in one or the other limits using the corresponding model.

\section*{Acknowledgments}
The authors are in debt to G. Endrodi for kindly sharing LQCD results for the critical temperature and condensate as functions of the field strength and to R. L. S. Farias for very useful conversations. The work was supported in part by UNAM-DGAPA-PAPIIT grant number IG100219 and by Consejo Nacional de Ciencia y Tecnolog\'ia grant numbers A1-S-7655 and A1-S-16215. M. L.  acknowledges support from Fondecyt (Chile) regular grants No. 1200483 and No. 1190192 and from
ANID/PIA/Basal (Chile) under grant FB082. C.V. acknowledges financial support from FONDECYT under grants 1190192 and 1200483.

\bibliographystyle{unsrt}
\bibliography{bibliography}

\end{document}